\documentclass[aps]{revtex4} 
\usepackage{psfig}
\usepackage{epsfig}
\usepackage{bm}

\newcommand{\be}{\begin{equation}}
\newcommand{\ee}{\end{equation}}

\begin{document} 
 
\title{Dynamics of ballistic annihilation} 
 
\author{Jaros\l aw Piasecki$\,^1$, Emmanuel Trizac$\,^2$ and Michel Droz$\,^3$}

\affiliation{
$^1$ Institute of Theoretical Physics, University of Warsaw, 
Hoza 69, 00-681 Warsaw, Poland \\
$^2$ Laboratoire de Physique 
Th{\'e}orique\footnote{Unit\'e Mixte de Recherche 8627 du CNRS}, 
B{\^a}timent 210, Universit{\'e} de Paris-Sud, 91405 Orsay, France \\
$^3$ D\'epartement de Physique, Universit\'e de Gen\`eve, 
CH-1211 Gen\`eve 4, Switzerland
}

\begin{abstract}
The problem of ballistically controlled annihilation is revisited
for general initial velocity
distributions and arbitrary dimension. 
An analytical derivation of the hierarchy equations obeyed by the reduced
distributions is given, and a scaling analysis of the corresponding
spatially homogeneous system is performed. This approach points
to the relevance of the non-linear Boltzmann equation 
for dimensions larger than one and provides
expressions for the exponents describing the decay
of the particle  density $n(t) \propto t^{-\xi}$ and the root mean-square
velocity ${\overline v} \propto t^{-\gamma}$ in term of a parameter
related to the dissipation of kinetic energy.  The Boltzmann equation is then
solved perturbatively within a systematic expansion in Sonine polynomials.
Analytical expressions for the exponents $\xi$ and $\gamma$ are
obtained in arbitrary dimension as a function of the parameter $\mu$
characterizing the small velocity behavior of the initial velocity
distribution. Moreover, the leading non-Gaussian corrections
to the scaled velocity distribution are computed.
These expressions for the scaling exponents 
are in good agreement with the values
reported in the literature  for continuous velocity distributions in
$d=1$.  For the two dimensional case, we implement Monte-Carlo
and molecular dynamics simulations that turn out to be 
in excellent agreement with the
analytical predictions. 
\end{abstract} 

\maketitle

%%%%%%%%%%%%%%%%%%%%%%%%%%%%%%%%%%%%%%%%%%%%%%%%%%%%%%%%%%%%%%%%%%%%%%%%%%%%%
\section{Introduction}

Ballistically controlled reactions provide simple examples of non-equilibrium systems
with complex kinetics. They consist of an assembly of particles with a given velocity 
distribution, moving freely between collisions in a $d$-dimensional space.
In the simplest version of these models we consider here,
when two particles meet, they instantaneously annihilate each other and disappear 
from the system. 
%Such reactions are relevant in growth and coarsening processes\cite{Krug,..}.
Despite its apparent simplicity, this problem is highly non trivial and has attracted 
substantial interest during the past years
\cite{Elskens,Piasecki,Droz,BenNaim1,BenNaim2,Rey,Kafri,Blythe,Krapivsky,Trizac}. 
The one-dimensional case where the particles can only have two velocities $\pm c$ has
been studied in a pioneering work by Elskens and Frisch \cite{Elskens}. 
In particular, they proved that for
a symmetric initial velocity distribution, the particle density is decreasing, in the
long time limit, as $n(t) \propto t^{-\xi} \propto t^{-1/2}$. 
The case of general distributions
in dimension $d=1$ was discussed by Piasecki \cite{Piasecki}, who reduced 
exactly the annihilation dynamics to a single closed equation for the two-particle 
conditional probability. Moreover, it was shown in the particular 
bimodal situation of a discrete velocity distribution ($\pm c$) that 
in one dimension, important dynamical
correlations are developing during the time evolution, invalidating mean-field or
Boltzmann-like approaches.
This exact approach was applied to the case of a three velocity distribution by 
Droz et al \cite{Droz}, with the result that the dynamical exponents were strongly 
depending upon the details of the initial velocity distribution.

No analytical solutions could be found for continuous velocity 
distributions. 
In one dimension, Ben-Naim et al.\cite{BenNaim1,BenNaim2} 
have shown that the exponent $\xi$ could depend on the 
behavior near the
origin of the initial velocity distribution. This problem has been revisited 
by Rey et al \cite{Rey}. Based on the exact theoretical 
approach \cite{Piasecki,Droz}, a
dynamical scaling theory was derived and its validity supported by
numerical simulations for several velocity distributions. This leads to the
conjecture that all continuous velocity distributions $f(v)$ that are symmetric, 
regular, and such that $f(0) \not = 0$ are attracted, in the long-time regime, 
towards the same distribution and thus belong to the same universality class. 
This conjecture was reinforced by numerical simulations in two dimensions
\cite{Trizac}.

The case of a continuous velocity distribution has also been approached recently by
Krapivsky and Sire \cite{Krapivsky}. Starting from a Boltzmann equation, they investigated
the decay of the particle density $n(t) \sim t^{-\xi}$ and the root mean-square 
velocity
${\overline v} \propto t^{-\gamma}$. They derived 
upper and lower bounds for the exponents as 
well as their leading expansion in $1/d$, valid in high dimension.
The main question with such an approach concerns the validity of a Boltzmann
equation. This is not justified in 1D and remains an open problem 
in higher dimensions.

The purpose of this paper is to give a first principle answer 
to this type of question.
The article is organized as follows. In section \ref{sec:analytical}, 
an original  
analytical derivation of the
equations governing the dynamics of ballistic annihilation is given. 
The hierarchy equations
obeyed by  the reduced distributions are obtained. It is shown that in the Grad limit,
the hierarchy formally reduces to a Boltzmann-like form for $d>1$. If the initial
reduced distributions factorize, the whole hierarchy reduces to one non-linear equation.
In the second part of section \ref{sec:analytical}, 
a scaling analysis  of the exact spatially homogeneous 
hierarchy is performed.
The exponents $\xi$ and $\gamma$ are showed to depend only on one parameter $\alpha$
related to the dissipation of energy. This scaling analysis turns out to be invalid for
the case $d=1$ with discrete velocity distributions, but correct in the
continuous case. Strong arguments are given
in favor of the validity of the Boltzmann approach for the case $d>1$ in the long time
limit.
The Boltzmann equation is then solved within a systematic approximation based on an
expansion in Sonine polynomials (section \ref{sec:Boltzmann}). 
The first non-Gaussian corrections to the scaled velocity distribution are computed and
predictions for the exponents $\xi$ and $\gamma$ are explicitly
worked out as a function of the dimension $d$ and the parameter 
$\mu$ characterizing the small velocity behaviour
of the initial velocity distribution: [$f({\bf v},t=0) \propto |{\bf v}|^\mu$ for 
$|{\bf v}| \to 0$]. These predictions for $\xi$ and $\gamma$ 
are asymptotically exact for large dimensions, and reproduce the
$1/d$ correction to the mean-field values. 
In 1D, they are in very good agreement with the exponents
reported in the literature \cite{Krapivsky} at the Boltzmann level.
In 2D, we implement extensive Direct Simulation Monte Carlo methods (DSMC) where
the non-linear Boltzmann equation is solved, and Molecular Dynamics (MD) simulations
where the exact equations of motion are integrated (section \ref{sec:numerical}). 
The agreement between the MD  
and DSMC routes confirms the validity of the Boltzmann approach, and the 
decay exponents measured are in exceptionally good 
agreement with the Sonine prediction.
Conclusions are drawn in section \ref{sec:concl}. 
A preliminary account of part of the results presented here has been published
elsewhere \cite{Trizac}.

%%%%%%%%%%%%%%%%%%%%%%%%%%%%%%%%%%%%%%%%%%%%%%%%%%%%%%%%%%%%%%%%%%%%%%%%%%%%%
\section{Exact results}
\label{sec:analytical}
\subsection{Derivation of the hierarchy}

Let $\Omega$ be a region of finite measure in $R^{2d}$. We denote by 
\begin{equation}
\mu_{k}^{\Omega}({\bf r}_{1},{\bf v}_{1},\; ...\; ,{\bf r}_{k},{\bf v}_{k};t)
\label{0}
\end{equation}  
the probability density for finding at time $t$ exactly $k$ particles
within $\Omega $ in the states $({\bf r}_{j},{\bf v}_{j})\in \Omega ,
\;\;\; j=1,2, ... ,k$, where ${\bf r}_{j}$ and ${\bf v}_{j}$ are the
position and the velocity vectors, respectively. The knowledge of the
densities $ \mu_{k}^{\Omega}$ for all $\Omega \in R^{2d}$ and 
$k=0,1,2, ...$, defines entirely the state of the system. For a given
region $\Omega$ the normalization condition reads
\begin{equation}
\mu_{0}^{\Omega}(t)+\sum_{k=1}^{\infty}\int_{\Omega}d{\bf r}_{1}
d{\bf v}_{1}\; ...\; \int_{\Omega}d{\bf r}_{k}d{\bf v}_{k}\;
\mu_{k}^{\Omega} ({\bf r}_{1},{\bf v}_{1},\; ...\; ,{\bf r}_{k},
{\bf v}_{k};t) = 1   \label{1}
\end{equation}
where $\mu_{0}^{\Omega}(t)$ is the probability of finding the region $\Omega$
void of particles at time $t$.

A necessary condition for the occurrence of a pair of particles at the
phase space points   $({\bf r}_{j},{\bf v}_{j}),\; ({\bf r}_{i},{\bf v}_{i})$
at time $t>0$ is that ${\bf r}_{ij}={\bf r}_{i}-{\bf r}_{j},\; 
{\bf v}_{ij}={\bf v}_{i}-{\bf v}_{j}$
belong to the region of the phase space with the characteristic function
\begin{equation}
\chi ({\bf r}_{ij},{\bf v}_{ij};t) = \theta (|{\bf r}_{ij}|-\sigma )
\left\{ 1-\theta ({\bf r}_{ij}\cdot {\bf v}_{ij})
\theta \left(\sigma - \sqrt{|{\bf r}_{ij}|^{2}-({\bf r}_{ij}\cdot 
\widehat{\bf v}_{ij})^{2}}\right) \right.
\label{2}
\end{equation}
\[\left. \times  \theta \left(|{\bf v}_{ij}|t-{\bf r}_{ij}\cdot \widehat{\bf
v}_{ij}+\sqrt{\sigma^{2}-|{\bf r}_{ij}|^{2}+({\bf r}_{ij}\cdot 
\widehat{\bf v}_{ij})^{2}} \right) \right\} \]
where $\sigma$ denotes the particle diameter, 
$\widehat{\bf v}_{ij}$ is a unit vector in the direction of the
relative velocity, and $\theta$ denotes the Heaviside distribution.
Indeed, moving backward in time the particles collide during the time
interval $(0,t)$ if and only if the following three conditions are
simultaneously satisfied
$$
\begin{array}{rlll}
\hbox{(i) }   & {\bf r}_{ij}\cdot {\bf v}_{ij}>0 & 
\hbox{ particles approach each other}   &  \\
\hbox{(ii) }   & \sigma > \sqrt{|{\bf r}_{ij}|^{2}-({\bf r}_{ij}\cdot 
\widehat{\bf v}_{ij})^{2}} &  \hbox{ the impact parameter is smaller than }
\sigma & \\
\hbox{(iii) } & |{\bf v}_{ij}|t > {\bf r}_{ij}\cdot \widehat{\bf
v}_{ij}-\sqrt{\sigma^{2}-|{\bf r}_{ij}|^{2}+({\bf r}_{ij}\cdot 
\widehat{\bf v}_{ij})^{2}} &
\hbox{ the time $t$ is long enough} & \\
& & \hbox{ for the overlapping configuration to occur} &
\end{array}
$$
Hence, $\chi ({\bf r}_{ij},{\bf v}_{ij};t)=1$, if and only if no
overlapping takes place during the time interval $(0,t)$.

At time $t$ particles $1,2,\; ... \; k$ occupy in $\Omega$ 
the one-particles  states
\begin{equation}
 ({\bf r}_{1},{\bf v}_{1}),({\bf r}_{2},{\bf v}_{2})
\; ... \; ({\bf r}_{k},{\bf v}_{k})    \label{3}
\end{equation}
with probability density (\ref{0}).
Using the characteristic function (\ref{2}) we can construct the
probability density for finding the same particles in the phase space
configuration
\begin{equation}
({\bf r}_{1}+{\bf v}_{1}dt,{\bf v}_{1}),({\bf r}_{2}+{\bf v}_{2}dt,
{\bf v}_{2})
\; ... \; ({\bf r}_{k}+{\bf v}_{k}dt,{\bf v}_{k})    \label{4}
\end{equation}
at time $(t+dt),\; dt>0$. It reads
\begin{equation}
\left[\prod_{i<j}^{k}\chi ({\bf r}_{ij}+{\bf v}_{ij}dt,
{\bf v}_{ij};t+dt)\right]
\mu_{k}^{\Omega}({\bf r}_{1},{\bf v}_{1},\; ...\; ,{\bf r}_{k},{\bf v}_{k};t) 
\label{5}
\end{equation}  
In the limit $dt \to 0^+$ the above expression takes the asymptotic form
\begin{equation}\label{6}
\mu_{k}^{\Omega}({\bf r}_{1},{\bf v}_{1},\; ...\; ,{\bf r}_{k},
{\bf v}_{k};t)
 \left[ 1+\sum_{i<}^{k}\sum_{j}^{k}\left(\frac{\partial}{\partial t}
+{\bf v}_{ij}\cdot \frac{\partial}{\partial {\bf r}_{ij} } \right)
\chi ({\bf r}_{ij},{\bf v}_{ij};t)dt \right]  
\end{equation}
Using the definition (\ref{2}) we find
\begin{equation}
\left(\frac{\partial}{\partial t}
+  {\bf v}_{ij}\cdot \frac{\partial}{\partial {\bf r}_{ij} } \right)
\chi ({\bf r}_{ij},{\bf v}_{ij};t)  
 =  (\widehat{\bf r}_{ij}\cdot {\bf v}_{ij})\delta (|{\bf r}_{ij}|
-\sigma ) [1 - \theta ({\bf r}_{ij}\cdot {\bf v}_{ij}) ]
\label{7}
\end{equation}
where $\widehat{\bf r}_{ij}={\bf r}_{ij}/|{\bf r}_{ij}|$.
We denote by $T^{v}(i,j)$ the right hand side of (\ref{7}) and rewrite
it in the form
\begin{equation}
T^{v}(i,j) = \sigma^{d-1}
\int\, d\widehat{\bm\sigma}(\widehat{\bm\sigma}\cdot {\bf v}_{ij})
\theta (-\widehat{\bm\sigma}\cdot {\bf v}_{ij})\delta ({\bf r}_{ij}-
\sigma \widehat{\bf \sigma})   \label{8}
\end{equation}
Here $\widehat{\bm\sigma}$ is the unit vector along the line passing
through the centers of the spheres at contact.
The integration with respect to the measure $d\widehat{\bm\sigma}$ is thus
the angular integration over the solid angle in $d$-dimensional space. 
The $\theta$ function in (\ref{8}) restricts this angular integral to the
hemisphere  corresponding to pre-collisional configurations.

Our aim is to construct the probability density $\mu ^{\Omega}_{k}$ at 
time $(t+dt)$ for $dt\to 0^+ $ :
\begin{equation}
\mu ^{\Omega}_{k}({\bf r}_{1}+{\bf v}_{1}dt,{\bf v}_{1},
\; ... \; ,{\bf r}_{k}+{\bf v}_{k}dt,{\bf v}_{k};t+dt)
= \mu ^{\Omega}_{k}({\bf r}_{1},{\bf v}_{1},
\; ... \; {\bf r}_{k},{\bf v}_{k};t)       \label{9}
\end{equation}
\[ + \left(\frac{\partial}{\partial t}
+\sum_{j=1}^{k}{\bf v}_{j}\cdot \frac{\partial}{\partial {\bf r}_{j} }
\right)
\mu ^{\Omega}_{k}({\bf r}_{1},{\bf v}_{1},\; ... \; {\bf r}_{k},
{\bf v}_{k};t)dt  \]
To this end we have still to add to the term (\ref{6}) the probability 
weights of two events. The first corresponds to the presence at time $t$ of
 $(k+2)$ particles within $\Omega $ in the states
\begin{equation}
({\bf r}_{1},{\bf v}_{1}),({\bf r}_{2},{\bf v}_{2}),
\; ... \; ,({\bf r}_{k},{\bf v}_{k}),({\bf r}_{+1},{\bf v}_{k+1}),
({\bf r}_{k+2},{\bf v}_{k+2})   \label{10}
\end{equation}
The state (\ref{10}) is then transformed into (\ref{3}) at time
$(t+dt)$ as the result of an annihilating collision between the particles
$(k+1)$ and $(k+2)$, during the time interval $(t,t+dt)$.
According to equations (\ref{6}) and (\ref{7}), the rate of the
occurrence of binary collisions between pairs $(i,j)$ is obtained by
applying the operator $[-T^{v}(i,j)]$ defined in (\ref{8}) to the
corresponding distribution. Hence, when $dt\to 0^+$, the $(k+1,k+2)$
annihilation process contributes to the density (\ref{9}) the term
\begin{equation}
- \int_{\Omega}d(k+1)\int_{\Omega}d(k+2)T^{v}(k+1,k+2)
\mu ^{\Omega}_{k+2}(1,2,\; ...\; ,k,k+1,k+2;t)dt \label{11}
\end{equation}
where the shorthand notation 
$ dj \equiv d{\bf r}_{j} d{\bf v}_{j}$ for $j=1,2, ... $ has been used.

Finally, we have to take into account the effects of the free flow of 
particles across the boundary $\partial\Omega$ of the region $\Omega$.
Indeed, the $k$-particle state can be created or destroyed by an
additional particle $(k+1)$ leaving or entering the considered region.
Denoting by $\widehat{\bf n}$ the unit vector normal to  $\partial\Omega$
oriented outwards, we get the term
\begin{equation}
\int d{\bf v}_{k+1}\int_{\partial\Omega}dS (\widehat{\bf n}\cdot {\bf v}_{k+1})
\mu ^{\Omega}_{k+1}(1,\; ...\; ,k,k+1;t)dt 
\label{12} 
\end{equation}
\[ = \int_{\Omega}d(k+1){\bf v}_{k+1}\cdot \frac{\partial}
{\partial {\bf r}_{k+1}}
\mu ^{\Omega}_{k+1}(1,\; ...\; ,k,k+1;t)dt \]
Here $dS$ is the measure of the surface area, and the equality
(\ref{12}) follows from Gauss' theorem. 

The enumerated events combine together to create the complete rate 
of change of the
probability density $\mu_{k}^{\Omega}$. As equivalent events have the
same probability measure we  can equate  (\ref{9}) with the sum of 
contributions (\ref{6}),(\ref{11}) and (\ref{12}) obtaining thus  the
hierarchy equations $( k = 1,2, ... )$
\begin{equation}\label{13}
\left(\frac{\partial}{\partial t}
+\sum_{j=1}^{k}{\bf v}_{j}\cdot \frac{\partial}{\partial {\bf r}_{j} }
- \sum_{i<}^{k}\sum_{j}^{k}T^{v}(i,j)\right)
\mu ^{\Omega}_{k}(1,\; ...\; ,k;t) =  
\ee
$$
 - \int_{\Omega}d(k\! +\! 1)\int_{\Omega}d(k\! +\! 2) 
T^{v}(k\! +\! 1,k\! +\! 2)
\mu ^{\Omega}_{k+2}(1,2,\; ...\;,k\! +\! 2;t)  
+ \int_{\Omega}d(k\! +\! 1)\, {\bf v}_{k+1}\cdot \frac{\partial}
{\partial {\bf r}_{k+1}}\, 
\mu ^{\Omega}_{k+1}(1,\; ...\; ,k,k\! +\! 1;t).
$$
Finally, the evolution equation for $\mu^{\Omega}_{0}$ follows from the
normalization condition (\ref{1}). This completes the derivation of the
infinite hierarchy of equations satisfied by the probability densities
$\mu_{k}^{\Omega}$.

From (\ref{13}) one can derive in a straightforward way the hierarchy
satisfied by the reduced distributions $f_{k}(1,2,\; ...\; ,k;t)$. They
are
relevant for the evaluation of physical parameters, as 
 $f_{k}(1,2,\; ...\; ,k;t)d1\; ... \; dk$ 
represents the measure of the number of  $k$-particle phase space
configurations, with $k$ particles occupying the one-particle states
$ ({\bf r}_{1},{\bf v}_{1}),({\bf r}_{2},{\bf v}_{2})\; ... \; 
({\bf r}_{k},{\bf v}_{k})$  at time $t$. The distributions $f_{k}$ are
related to the probability densities $\mu_{k}^{\Omega}$ by the equation
\cite{Ruelle}
\begin{equation}\label{14}
f_{k}(1,2,\; ...\; ,k;t) = 
\sum_{p=0}^{\infty}\frac{(k\! +\! p)!}{p!}
\int_{\Omega}d(k\! +\! 1)...\int_{\Omega}d(k\! +\! p)
\mu^{\Omega}_{k+p}(1,\; ...\; ,k,k\! +\! 1,\; ...\; ,k\! +\! p;t).
\end{equation}
Note that the $f_{k}(1,2,\; ...\; ,k;t)$ do not depend on $\Omega$. 

In order to derive the evolution equation for $f_{k}$ one has thus to consider
the hierarchy equation (\ref{13}) with $k$ replaced by $(k+p)$, and use
the relation (\ref{14}). One finds
\begin{equation}\label{15}
\left(\frac{\partial}{\partial t}
+\sum_{j=1}^{k}{\bf v}_{j}\cdot \frac{\partial}{\partial {\bf r}_{j} }
- \sum_{i<}^{k}\sum_{j}^{k}T^{v}(i,j)\right)
f_{k}(1,\; ...\; ,k;t)             
 =  \sum_{p=0}^{\infty}\frac{(k\! +\! p)!}{p!}
\int_{\Omega}d(k\! +\! 1)...\int_{\Omega}d(k\! +\! p) 
\end{equation}
\[ \left\{  \left[- \sum_{j=k+1}^{k+p}
{\bf v}_{j}\cdot \frac{\partial}{\partial {\bf r}_{j}} 
+ \sum_{j=1}^{k}\, \sum_{i=k+1}^{k+p}T^{v}(i,j)
+ \sum_{k+1\leq i}^{k+p}\sum_{<j}^{k+p}T^{v}(i,j) \right]
\mu_{k+p}(1,\; ...\; ,k\! +\! p;t) \right. \]
\[ \left. - \int_{\Omega}d(k\! +\! p\! +\! 1)\int_{\Omega}d(k\! +\! p\! +\!2)
T^{v}(k\! +\! p\! +\! 1,k\! +\! p\! +\! 2)
\mu_{k+p+2}(1,\; ...\; ,k\! +\! p\! +\! 1,k\! +\! p\! +\! 2;t) \right. \]
\[\left. + \int_{\Omega}d(k\! +\! p\! +\! 1)\left({\bf v}_{k\! +\! p\! +\! 1}
\cdot \frac{\partial}{\partial {\bf r}_{k+p+1}}\right) 
\mu ^{\Omega}_{k+p+1}(1,\; ...\; k\! +\! p\! +\! 1;t) \right\} \]

It is then a question of inspection to see that on the right hand side
of (\ref{15}) only the term
\begin{equation}
\sum_{p=1}^{\infty}\frac{(k+p)!}{p!}\int_{\Omega}d(k+1)...\int_{\Omega}d(k+p)
\sum_{j=1}^{k}\sum_{i=k+1}^{k+p}T^{v}(i,j)
\mu_{k+p}(1,\; ...\; ,k\! +\! p;t)  \label{16}
\end{equation}
\[ = \int d(k\! +\! 1)\sum_{j=1}^{k}T^{v}(j,k\! +\! 1)f_{k+1}
(1,\; ...\; ,k,k\! +\! 1;t)  \]
survives. All the remaining terms exactly  cancel out.

The hierarchy equations satisfied by the reduced distributions $f_{k}$
 describing the annihilation dynamics thus read
\begin{equation}\label{eq:hierarchy}
\left(\frac{\partial}{\partial t}
+\sum_{j=1}^{k}{\bf v}_{j}\cdot \frac{\partial}{\partial {\bf r}_{j} }
- \sum_{i<}^{k}\sum_{j}^{k}T^{v}(i,j)\right)
f_{k}(1,\; ...\; ,k;t) 
 = \int d(k\! +\! 1)\sum_{j=1}^{k}T^{v}(j,k\! +\! 1)f_{k+1}
(1,\; ...\; ,k,k\! +\! 1;t).
\end{equation}
Consider now equations (\ref{eq:hierarchy}) supposing that the state of the
system is spatially homogeneous. In this case the distribution $f_{1}$
does not depend on the particle position. Let us formally take the Grad
limit
\begin{equation}
\sigma \to 0,\;\;\; n(t) \to \infty, \;\;\;  n(t)\sigma^{d-1} =
 \lambda^{-1} = const 
\label{18}
\end{equation}
where
 \[  n(t) = \int d{\bf v}\; f_{1}({\bf v};t)\] 
The fixed mean free path $\lambda$ introduces a relevant length scale,
so we pass to dimensionless positions putting 
\begin{equation}
{\bf r}_{j}=\lambda {\bf x}_{j},\;\;\; j=1,2, ...         \label{19} 
\end{equation} 
With this change of variables the collision operator (\ref{8}) takes the
form
\begin{equation}
T^{v}(i,j) = [n(t)\sigma^{d}]^{d-1}\frac{1}{\lambda}
\int\, d\widehat{\bm\sigma}(\widehat{\bm\sigma}\cdot {\bf v}_{ij})
\theta (-\widehat{\bm\sigma}\cdot {\bf v}_{ij})\delta [{\bf x}_{ij}-
 n(t)\sigma^{d} \widehat{\bf \sigma}] 
\label{20}
\end{equation} 
We conclude that the term on the left hand side of equation (\ref{eq:hierarchy})
involving the collision operators (\ref{20}) vanishes in the Grad limit for $d>1$,
because the dimensionless parameter $n(t)\sigma^{d}$ tends to zero.
Note that this term induces dynamical correlations, hindering the propagation
of the molecular chaos factorization.
On the other hand, using the definition of $T^{v}(j,k\! + \!1)$ we find
that the term on the right hand side equals
\begin{equation} \label{21}
\frac{1}{\lambda}\sum_{j=1}^{k}\int d{\bf v}_{k+1}\int\, d\widehat{\bm\sigma}
(\widehat{\bm\sigma}\cdot {\bf v}_{j (k\! +\! 1)})
\theta (-\widehat{\bm\sigma}\cdot {\bf v}_{j (k\! +\! 1)})   
f_{k+1}(1,\; ...\; ,j, \; ...\; ,k,
{\bf r}_{k+1}={\bf r}_{j}\! -\! \sigma\widehat{\bf \sigma},
{\bf v}_{k+1};t)/n(t) 
\end{equation}
and its prefactor $1/\lambda$ remains finite in the same limit. 
So, formally the hierarchy equations (\ref{eq:hierarchy}) at a given time $t$
reduce pointwise in the Grad limit to the Boltzmann-like hierarchy
\begin{equation}\label{22}
\left(\frac{\partial}{\partial t}
+\sum_{j=1}^{k}{\bf v}_{j}\cdot \frac{\partial}{\partial {\bf r}_{j} }
\right)f^{B}_{k}(1,\; ...\; ,k;t) 
 = \int d(k\! +\! 1)\sum_{j=1}^{k}T^{v}(j,k\! +\! 1)f^{B}_{k+1}
(1,\; ...\; ,k,k\! +\! 1;t),\;\;\;  (k = 1,2,\; ...) 
\end{equation}

The hierarchy (\ref{22}) propagates the factorization of the
reduced distributions
\begin{equation}
f^{B}_{k}(1,\; ...\; ,k;t)=\prod_{j=1}^{k}f^{B}_{k}(j;t)   \label{23}
\end{equation}
Hence, if the initial state is factorized, the whole hierarchy (\ref{22})
reduces to one  non-linear equation 
\begin{equation}
\left(\frac{\partial}{\partial t}
+{\bf v}_{1}\cdot \frac{\partial}{\partial {\bf r}_{1}}
\right)f^{B}(1;t)= \int d2\; T^{v}(1,2)f^{B}(1;t)f^{B}(2;t)
\label{24}
\end{equation}  
Equation (\ref{24}) is the Boltzmann kinetic equation corresponding
to the annihilation dynamics. In the following section, we shall see
that the formal Grad limit taken here, where $n\to \infty$,
is relevant for the description 
of the annihilation dynamics at late times, even if the density $n(t)$
decreases with time.

%%%%%%%%%%%%%%%%%%%%%%%%%%%%%%%%%%%%%%
\subsection{Scaling analysis of the hierarchy}

The evolution of the annihilation kinetics shares a common feature 
with the Grad limit: The ratio of particle diameter to mean-free-path 
$\lambda = 1/(n\sigma^{d-1})$,
which is related to the packing fraction, vanishes in both cases. To be more 
specific, we perform a scaling analysis of the exact {\em homogeneous}
hierarchy equations
and look for self-similar reduced distributions where the time dependence
has been absorbed into the density $n(t)$, with the velocities ${\bf v}$
and positions ${\bf r}$ renormalized by the typical (root mean squared) 
velocity $\overline v (t)$
and mean-free-path respectively: We define the reduced variables 
\be
{\bf c} = \frac{{\bf v}}{\overline v } \qquad \hbox{and} \qquad 
{\bf x} = \frac{\bf r}{\lambda}, \qquad \hbox{with}\qquad
(\overline v) ^2 = \frac{1}{n(t)}\, \int v^2 f_1({\bf v};t)\,d{\bf v}.
\label{eq:defrescaled}
\ee
For the one-body distribution, we therefore introduce the 
reduced function $\widetilde f_1$ such that
\be
f_1({\bf v};t) \, = \, \frac{n(t)}{\overline v (t)^d}\,\, \widetilde f_1({\bf c})
 \, = \, \frac{n(t)}{\overline v (t)^d}\,\, 
\widetilde f_1\left( \frac{{\bf v}}{\overline v(t) } \right).
\label{eq:scaling1}
\ee
By definition, the moments of order 0 and 2 of $\widetilde f_1$ 
are constrained to unity.
Requiring that the $k$-body distribution $f_k$ factorizes into
$\prod_{i=1}^k f_1(i)$ in the limit
of infinite relative separations between the particles, we consistently obtain
the scaling form:
\be
f_k({\bf r}_1,{\bf v}_1, ..., {\bf r}_k,{\bf v}_k;t) = 
\left( \frac{n}{\overline v ^d}\right)^k \widetilde f_k({\bf x}_1,{\bf c}_1, ...,
{\bf x}_k,{\bf c}_k). 
\label{eq:scaling2}
\ee

The evolution equations of $n(t)$ and kinetic energy density 
$n \overline v ^2(t)$ follow from integrating 
(\ref{eq:hierarchy}) with weights $d {\bf v}_1$ and $ v_1^2 d{\bf v}_1$ 
respectively, for $k=1$. We obtain
\begin{eqnarray}
&&\frac{d n}{dt} = - \omega(t)\, n
\label{eq:alphadef1}\\
&&\frac{d (n \overline v ^2)}{dt} =
% -\omega(t) n T_{\hbox{\scriptsize coll}} =
  - \alpha\, \omega(t)\, n\, \overline v ^2,
\label{eq:alphadef2}
\end{eqnarray}
where the collision frequency $\omega$ and kinetic energy dissipation parameter
$\alpha$ read
\begin{eqnarray}
&&\omega(t) \,=\, n(t) \overline v (t)\, 
\int d{\bf c}_1d{\bf c}_2 d \widehat{\bm \sigma}\,
(-\widehat{\bm \sigma}\cdot{\bf c}_{12})\, 
\theta(-\widehat{\bm \sigma}\cdot{\bf c}_{12})\, \widetilde f_2({\bf c}_1,{\bf c}_2,
\sigma \widehat{\bm \sigma}) 
\label{eq:omegadef}\\
&& \alpha = \frac{\int d{\bf c}_1d{\bf c}_2 d \widehat{\bm \sigma}\,
(\widehat{\bm \sigma}\cdot{\bf c}_{12})\, 
\theta(-\widehat{\bm \sigma}\cdot{\bf c}_{12})\, c_1^2\,
\widetilde f_2({\bf c}_1,{\bf c}_2, \sigma \widehat{\bm \sigma})}
{\left[\int c^2 \widetilde f_1({\bf c})\,d{\bf c} \right] \, \left[
\int d{\bf c}_1d{\bf c}_2 d \widehat{\bm \sigma}\,
(\widehat{\bm \sigma}\cdot{\bf c}_{12})\, 
\theta(-\widehat{\bm \sigma}\cdot{\bf c}_{12})\,
\widetilde f_2({\bf c}_1,{\bf c}_2, \sigma \widehat{\bm \sigma})\right]}.
\label{eq:alpha32}
\end{eqnarray}
Equation (\ref{eq:alpha32}) is valid for general velocity rescalings;
The definition (\ref{eq:defrescaled}) chosen here implies that
the term $\int c^2 \widetilde f_1({\bf c})\,d{\bf c}$ in the denominator
equals unity.
The coefficient $\alpha$ may be seen as the ratio of the kinetic energy 
dissipated in a typical collision normalized by the average kinetic energy,
and is time-independent in the scaling regime. 
%We therefore expect $\alpha>1$.
It is convenient to introduce
the internal ``clock'' ${\cal C} $ of the dynamics counting the number of collisions,
such that $d{\cal C} = \omega dt$. With this variable, Eqs. (\ref{eq:alphadef1}) 
and (\ref{eq:alphadef2}) integrate into
\be
n(t) \,=\, n_0\, \exp[-{\cal C}(t)] \qquad \hbox{and} \qquad 
\overline v^2(t) \,=\, \overline v^2_0\,\exp[-(\alpha-1) {\cal C}(t)],
\label{eq:ndeC}
\ee
where the time origin with density $n_0$ and kinetic energy density
$n_0 \overline v^2_0$ has been chosen to coincide with 
${\cal C}=0$. Knowledge of the ${\cal C}$-dependence of $n$ and 
$\overline v$ allows to relate absolute time $t$ to the number of
accumulated collisions: From Eq. (\ref{eq:omegadef}), we have
\be
\frac{d {\cal C}}{d t} \,=\, \omega_0 \frac{n}{n_0}\,\frac{\overline v}{\overline v_0}
\,= \, \omega_0 \,\exp[-{\cal C}(1+\alpha)/2],
\ee
where $\omega_0 \equiv \omega(t=0)$, so that 
\be
{\cal C} \,=\, \frac{2}{1+\alpha}\ln\left(1+ \frac{1+\alpha}{2} \,\omega_0 t
\label{eq:calC}
\right).
\ee
The corresponding time evolution is
\begin{eqnarray}
&& \frac{n}{n_0}\,=\, \left[1 + \frac{1+\alpha}{2}\,\, \omega_0 t
\right]^{-\frac{2}{1+\alpha}} 
\label{eq:timen}\\
&& \frac{\overline v}{\overline v_0}\,=\, \left[1 + \frac{1+\alpha}{2}\,\, \omega_0 t
\right]^{\frac{1-\alpha}{1+\alpha}}.
\label{eq:timev}
\end{eqnarray}
Without knowing the detailed form of the one-particle distribution function 
$f_1$, it is thus possible to conclude about the time 
decay of $n$ and $\overline v$,
which appear to obey algebraic laws in the long time limit
[$n(t) \propto t^{-\xi}$ and $\overline v(t) \propto t^{-\gamma}$,
with $\xi = 2/(1+\alpha)$ and $\gamma = (\alpha-1)/(\alpha+1)$]. 
The exponents $\xi$ and $\gamma$ are consequently simply related to 
the unknown quantity $\alpha$, for which a perturbative expansion will be
put forward in section \ref{sec:Boltzmann} before a numerical investigation
in section \ref{sec:numerical}. 
Moreover, if the initial velocity 
distribution is of finite support (i.e. vanishes outside a sphere
of given velocity $v^\star$), $\overline v$ fulfills the bound
$\overline v \leq v^\star$ so that $\gamma$ is necessarily positive
(or $\alpha \geq 1$). In the framework of Boltzmann's equation,
it will be shown in appendix \ref{app:alpha} that the quantity
$\alpha$ is necessarily larger than 1.
For the specific initial condition
where all particles have the same kinetic energy at $t=0$, $\overline v^2=
\langle v^2 \rangle$ is time-independent, and
(\ref{eq:timev}) implies that $\alpha=1$. From Eq. (\ref{eq:timen}),
we therefore obtain the time evolution for this situation
\be
\frac{n}{n_0}\,=\, \frac{1}{1 + \omega_0 t},
\label{eq:solpart}
\ee
which is exact within the scaling theory. This relation {\it a priori}
holds in any dimension, except for $d=1$ where the corresponding
initial velocity distribution is the symmetric discrete bimodal
function $\pm c$, for which the scaling ansatz underlying
our approach fails (see the discussion at the end of the present section).

Inserting the scaling forms (\ref{eq:scaling1}) and (\ref{eq:scaling2})
into the first equation of the hierarchy (\ref{eq:hierarchy}) imposes
the following constraint on the decay exponents: $\xi + \gamma=1$.
This scaling relation may be simply obtained by elementary dimensional analysis
\cite{BenNaim1,BenNaim2,Rey,Krapivsky,Trizac}, 
and may be considered as the compatibility
condition of the hierarchy with the self-similar scaling solutions
\cite{Trizac_coal}.
It is moreover identically fulfilled by the expressions
(\ref{eq:timen}) and (\ref{eq:timev}). 
Under the constraint $\xi+\gamma=1$, the remaining equations of the 
hierarchy ($k>1$) turn out to be compatible with (\ref{eq:scaling2})
with the additional information that the collision term on the lhs
of (\ref{eq:hierarchy}) decays like $t^{-\gamma -d \xi}$ whereas the 
remaining terms are associated with a power $1/t$. Given that
$\gamma + d \xi = 1 + (d-1) \xi \geq 1$, this collision term is asymptotically
irrelevant except in one dimension where it remains of the same
order as the dominant ones ($1/t$). We therefore recover the conclusions 
obtained by considering the formal Grad limit, with distribution 
functions expected to obey a Boltzmann-like equation.
This analysis points to the relevance of Boltzmann equation for $d>1$,
a point which is further corroborated by the numerical results
given in section \ref{sec:numerical}. 

It is interesting to note that both Eq. (\ref{eq:solpart}) 
and the relation $\xi+\gamma=1$ do not hold in 1D for discrete 
initial velocity distributions. In the symmetric situation 
of a bimodal distribution, the average kinetic energy per particle is 
conserved (so that $\gamma=0$), whereas the density decays as 
$1/\sqrt{t}$ (i.e. $\xi =1/2$ \cite{Elskens}). 
The scaling assumption (\ref{eq:scaling2}) is consequently
incorrect in the specific situation of discrete distributions
in 1D, but valid for continuous distributions \cite{Rey}.
To be more specific, the scaling form (\ref{eq:scaling2}) implies
that the collision frequency scales with time like $\omega \propto n \overline v$.
On the other hand, from the analytical solution of the bimodal 
$\pm c$ situation \cite{Elskens}, we obtain $\omega \propto n^2 \overline v$ with 
$\overline v = c$. This discrepancy is the signature of dynamical
correlations in the latter discrete case.
These correlations are responsible for the breakdown of (\ref{eq:scaling2}),
and in addition violate molecular chaos.
For continuous velocity distributions, again in 1D, molecular chaos
also breaks down while the scaling (\ref{eq:scaling2}) is correct.
As a consequence, the exponents obtained
at the Boltzmann level differ from the exact ones (see the discussion 
in the last paragraph of section \ref{sec:Boltzmann}), whereas
the relation $\xi+\gamma=1$ holds.

%%%%%%%%%%%%%%%%%%%%%%%%%%%%%%%%%%%%%%%%%%%%%%%%%%%%%%%%%%%%%%%%%%%%%%
\section{Boltzmann kinetic equation}
\label{sec:Boltzmann}

This section is devoted to the analysis of the decay
dynamics within the molecular chaos framework \cite{Resibois} of 
the homogeneous non-linear Boltzmann equation. No exact solution could be 
obtained, and our goal
is to derive accurate approximate predictions for the scaling 
exponents $\xi$ and $\gamma$ of the density and root mean squared velocity. 

Before considering the kinetic equation obeyed by the rescaled
distribution function, it is instructive to rewrite the original homogeneous 
Boltzmann equation (\ref{24}) in the form
\be
\frac{\partial f_1({\bf v};t)}{\partial t}\,=\,- \nu({\bf v};t) f_1({\bf v};t)
\quad \hbox{with} \quad 
\nu({\bf v_1};t) =  \left[\sigma^{d-1}
\int d\widehat{\bm\sigma}(\widehat{\bm\sigma}\cdot \widehat{\bf v}_{12})
\theta (\widehat{\bm\sigma}\cdot \widehat{\bf v}_{12})
\right] \,\int d{\bf v}_2 |{\bf v}_{12}| f_1({\bf v}_2;t)
\label{eq:boltznu}
\ee
where in the last equation, the term in brackets may be computed explicitly
as function of dimension (it is understood that the unit vector
$\widehat{\bf v}_{12}$ denotes an arbitrary direction).  
For our purpose, it is sufficient to notice
that at all times, the collision frequency $\nu({\bf v};t)$ 
of the population having velocity 
${\bf v}$ remains finite in the limit $v \to 0$, provided the 
first moment of $f_1$ exists. In this situation, Eq. (\ref{eq:boltznu})
implies that $f_1({\bf v};t)/f_1({\bf v};0)$ admits a finite limit
for $v\to 0$, or equivalently, it may be stated that if the initial 
velocity distribution behaves like $v^\mu$ near the velocity origin, 
this feature is preserved by the Boltzmann dynamics. Previous works have shown
accordingly that the scaling exponents $\xi$ and $\gamma$  
depend on the exponent $\mu$ \cite{BenNaim1,Rey,Krapivsky,Trizac}. 

Making use of relations
(\ref{eq:alphadef1}) and (\ref{eq:alphadef2}), 
insertion of the scaling form (\ref{eq:scaling1}) into
the Boltzmann equation leads to 
\be
\left[1 + 
\left(\frac{1-\alpha}{2}\right)\left(d+c_1\frac{d}{dc_1}\right)\right]\,
\widetilde f_1( c_1) \,=\,
\widetilde f_1(c_1) \int\! d{\bm c_2}\, 
\frac{c_{12}}{\langle c_{12}\rangle}\, \widetilde f_1 (c_2),
\label{eq:Boltzrescaled}
\ee
where we have assumed an isotropic velocity distribution 
[$\widetilde f_1({\bf c}) = \widetilde f_1(c) $]
and introduced the average 
$\langle (\ldots) \rangle = \int (\ldots) 
\widetilde f_1(c_1)\,\widetilde f_1(c_2)d{\bm c}_1 d{\bm c}_2$. 
$\langle c_{12}\rangle$ is therefore the rescaled collision frequency. 
Once $\mu$ has been chosen, Eq. (\ref{eq:Boltzrescaled}) admits
a solution for a unique value of $\alpha$. We show in appendix
\ref{app:alpha} that the inequality $\alpha >1$ necessarily holds.

Irrespective of $\alpha$, the large velocity behaviour of $\widetilde f_1$
may be obtained following similar lines as in 
\cite{BenNaim1,BenNaim2,Krapivsky,vanNoije,BBRTW}:
it is possible to integrate formally (\ref{eq:Boltzrescaled}) and cast 
$\widetilde f_1$ into
\be
\frac{\widetilde f_1(c)}{\widetilde f_1(c')}  \,=\, 
\left(\frac{c}{c'}\right)^{-\frac{2+d(1-\alpha)}{1-\alpha}} \, 
\exp\left[\frac{2}{1-\alpha}\,\frac{1}{\langle c_{12}\rangle} 
\int_{c'}^c \frac{\widetilde \nu(c'')}{c''} d c''
\right].
\label{eq:ratiof}
\ee
In this equation, $\widetilde \nu$ is itself a functional of $\widetilde f_1$:
\be
\widetilde \nu(c_1) \,=\, \int c_{12} \,\widetilde f_1(c_2)\, d {\bf c}_2,
\ee
such that $\widetilde \nu(c) / c $ goes to a finite limit for 
$c \to \infty$. We therefore obtain the large velocity tail 
\be
\widetilde f_1(c) \propto c^{-\frac{2+d(1-\alpha)}{1-\alpha}} \, 
\exp\left(-\frac{2}{\alpha-1}\frac{c}{\langle c_{12}\rangle}\right)
\qquad \hbox{for} \qquad c \to \infty.
\label{eq:tail}
\ee
In one dimension, we recover the results of references \cite{BenNaim1} and 
\cite{Krapivsky}. In \cite{BenNaim1,BenNaim2}, an approximation was derived 
for $\alpha$ [or equivalently $\xi = 2/(1+\alpha)$] by assuming that the 
large velocity behaviour of \smash{$\widetilde f_1$} could hold for the whole
velocity spectrum. In this picture, the power of $c$ appearing on the 
rhs of Eq. (\ref{eq:tail}) is equated to the exponent $\mu$ characteristic
of the small velocity behaviour (imposed by the initial distribution 
chosen, see above), with the result
\be
\alpha \,=\, 1 + \frac{2}{\mu + d} \qquad \hbox{or} \qquad 
\xi \,=\, \frac{2d + 2\mu}{2d + 2\mu +1}. 
\label{eq:large}
\ee
This prediction encodes the correct dependence on $\mu$ and
dimension ($\xi$ increases when $\mu$ or $d$ increase), and turns out 
to have an accuracy of order 10\% when compared to the numerical
results \cite{Trizac}. In the limit of large dimension, we obtain 
from (\ref{eq:large}) $\xi \sim 1-(2d)^{-1}$, whereas Krapivsky and Sire
have shown that $\xi \sim 1-d^{-1}(1-1/\sqrt{2})$, also in the framework of
the Boltzmann equation. The remainder of this section is devoted to the derivation 
of a more precise value for $\alpha$, which furthermore coincides with the exact
$1/d$ correction for $d \to \infty$.

Invoking the identity
\be
\int d {\bf c} \, c^p \left(d+c\frac{d}{dc}\right) \widetilde f_1({\bf c}) \,=\,
-p \, \langle c^p \rangle,
\ee
the energy dissipation parameter $\alpha$ may be given the set of 
equivalent expressions:
\be
\alpha \,=\, 1 \,+\,\frac{2}{p}\left(
\frac{\langle c_{12}\,c_1^p\rangle}{\langle c_{12}\rangle\langle c_1^p\rangle}-1
\right), \qquad \forall p \geq 0.
\label{eq:alphautile1}
\ee
A particularly useful relation between $\alpha$ and moments of $\widetilde f_1$
follows from considering the limit $c_1 \to 0$ of (\ref{eq:Boltzrescaled}):
we get 
\be 
\alpha = 1+ \frac{2}{\mu + d}\left(
1-\frac{\langle c_1 \rangle}{\langle c_{12}\rangle}\right). 
\label{eq:alphautile0}
\ee

The (infinite) family of relations (\ref{eq:alphautile0}) and (\ref{eq:alphautile1})
is equivalent to the original integro-differential equation (\ref{eq:Boltzrescaled}),
and well suited to a perturbative analysis. To this end, a systematic approximation 
of the isotropic function $\widetilde f_1$ can be found by expanding it in a set
of Sonine polynomials \cite{Landau}:
\be
\widetilde f_1(c) \, = \, {\cal M}(c)\, \left[
1 + \sum_{n=1}^\infty a_n\, S_n(c^2)    \right].
\label{eq:Sonine}
\ee
These polynomials are orthogonal with respect to the Gaussian weight 
\be
{\cal M}(c) = \left(\frac{d}{2\pi}\right)^{d/2} \, e^{-d c^2/2},
\ee
and the first few read
\begin{eqnarray}
&& S_0(x) \,=\, 1 \\
&& S_1(x) \,=\, \frac{d}{2} \left( -x + 1 \right)\\
&& S_2(x) \,=\, \frac{d^2}{8}\,x^2\,-\,\frac{d(d+2)}{4}\, x\,+\,\frac{d(d+2)}{8}.
\end{eqnarray}
The coefficients $a_n$ follow from the orthogonality relation 
$\int S_n(c^2) S_m(c^2) {\cal M}(c) d {\bf c} \propto \delta_{nm}$: In particular, 
\be
a_1 = \frac{2}{d}\langle S_1(c^2) \rangle = 
\frac{2}{d} \left(1-\langle c^2 \rangle\right) = 0
\ee
from the definition of rescaled velocities (\ref{eq:defrescaled}). The first
non-Gaussian correction is thus embodied in $a_2$, that is proportional to 
the fourth cumulant (kurtosis) of the velocity distribution: 
\be
a_2 \,=\, \frac{d^2}{3} \left[
\langle c_i^4\rangle -3\langle c_i^2 \rangle^2 \right] \,=\,
\frac{d}{d+2}\, \langle c^4\rangle \,-\, 1,
\label{eq:a2def}
\ee
with $c_i$ a Cartesian component of ${\bf c}$. 
Upon truncating Eq. (\ref{eq:Sonine}) at a finite order $n$, 
we obtain a regular velocity distribution near $c=0$. We consequently
restrict our analysis to the case $\mu=0$ (the dependence on $\mu$ 
has been considered in \cite{Trizac}). 

It is also noteworthy that any truncation 
of (\ref{eq:Sonine}) at arbitrary order $n$
leads to a Gaussian high energy behaviour, incompatible with the result 
(\ref{eq:tail}) corresponding to an overpopulated tail with respect to the
Maxwellian. However, it will be shown in section \ref{sec:numerical}
that the difference between the truncated
expansion (\ref{eq:Sonine}) and the numerical velocity distribution 
becomes manifest far in the tail, where the distribution has reached very low 
probabilities. Consequently, when the moment involved in (\ref{eq:alphautile1}) 
and (\ref{eq:alphautile0}) are evaluated from the truncation of 
(\ref{eq:Sonine}), the accuracy of the result is expected to be better 
for low orders $p$ in (\ref{eq:alphautile1}). Hence the privileged role played
by (\ref{eq:alphautile0}), which is of lower order than any of the identities
(\ref{eq:alphautile1}). In practice, upon truncating (\ref{eq:Sonine})
at order $n$, the $n$ unknowns $\alpha$, $a_2$, \ldots, $a_n$ are computed 
evaluating the corresponding moments appearing in (\ref{eq:alphautile0})
and in $n-1$ of the relations (\ref{eq:alphautile1}), among which it is 
convenient to retain the $n-1$ even values of $p$ ($p=0$ excluded). 
Truncation of (\ref{eq:Sonine}) at order $n=2$ yields precise predictions for 
$\alpha$ and $\widetilde f_1$, and already at Gaussian order, $\alpha$ turns
out to be very close to its numerical counterpart: Setting $n=0$ (or equivalently
$n=1$ since $a_1 \equiv 0$) in (\ref{eq:Sonine}), we obtain immediately the zeroth
order approximation 
\be
\alpha \,=\, \alpha_0 \,=\, 1 + \frac{2}{d}\left(1-\frac{\sqrt{2}}{2}\right)
\label{eq:alpha0}
\ee
which corresponds to 
\be
\xi \,=\, \xi_0 \,=\, \frac{2}{1+\alpha_0} \,=\, 
\frac{2 d }{2(d+1) -\sqrt{2}}.
\label{eq:xi0}
\ee
For large dimensions, this estimation goes to unity, with the 
exact $1/d$ correction computed in \cite{Krapivsky}
\be
\xi = 1- \frac{1}{d}\,\left( 1-\frac{\sqrt{2}}{2} \right) \,+\,
{\cal O}\left( \frac{1}{d^2}\right).
\label{eq:xilarged}
\ee
This behaviour turns out to be ``universal'', in the sense 
that the $\mu$ dependence does not appear at this order \cite{Trizac}.

We shall also be interested in the non-Gaussian features of the velocity distribution,
that we quantify by the fourth cumulant $a_2$. The (cumbersome) calculations at second
order in Sonine expansion are detailed Appendix \ref{app:Sonine}. We obtain 
\begin{eqnarray}
&& a_2 \, =  \,8 \,\frac{ d (2 \sqrt{2}-3)}
{ 4 d^2  + d (6  -\sqrt{2})} 
\label{eq:a2} \\
&& \alpha_2 \, = \, \alpha_0 \, + \, \frac{\sqrt{2}}{16 d} \, a_2. 
\label{eq:alpha2}
\end{eqnarray}
The corresponding density exponent follows from $\xi_2 = 1/(1+\alpha_2)$ as before.
For $d \to \infty$, $a_2 \sim 2 (3-2\sqrt{2}) d^{-1}$, irrespective of 
$\mu$ \cite{Trizac}
which reinforces the ``universal'' nature of large $d$. 
The correction to $\xi$ carried
by $a_2$  behaves as $1/d^2$ in this limit, and does not affect the $1/d$ terms 
which still coincide with the exact behaviour (\ref{eq:xilarged}).
Both predictions (\ref{eq:alpha0}) and (\ref{eq:alpha2}) are such that 
$\alpha >1$, which is required to obtain a normalizable distribution in
(\ref{eq:ratiof}) and (\ref{eq:tail}).

The second order expansion considered here may be improved 
by consideration of higher order Sonine terms and inclusion of
non-linear terms in $a_2$. In the related context of 
inelastic hard spheres, the limitation of working at linear order
in $a_2$ with neglect of Sonine terms of order $n \geq 3$
may be found in \cite{Santos}. Alternatively, keeping
non linear terms in $a_2$ and neglecting again Sonine terms
of order $n\geq 3$ leads to multiple solutions. A stability 
analysis is then required to determine which one
is stable, as discussed in \cite{Brilliantov}.

Here, the value obtained for $a_2$ is quite small (see below).
Our approximate expressions are accurate when compared to the full
numerical solution of Boltzmann's equation, so that 
we did not calculate any higher order coefficients, nor 
consider non linear terms in $a_2$.  
The existing body of literature 
reports, within Boltzmann framework, numerical exponents in 1D that are 
in excellent agreement with our predictions, already at zeroth order. 
For the case $\mu=0$, expressions
(\ref{eq:xi0}) and (\ref{eq:alpha2}) give $\xi_0 \simeq 0.773$ 
and $\xi_2 \simeq 0.769$
whereas the numerical result obtained in \cite{Krapivsky} is
$\xi \simeq 0.769$. These exponents are close to their
counterparts extracted numerically from the exact dynamics 
(0.785 $\pm 0.005$ in \cite{Rey}, and more recently 0.804 \cite{Com}).
The difference between the exact exponents and those obtained assuming 
molecular chaos is consistent with the conclusion of section 
\ref{sec:analytical}: 
In 1 dimension, the factorization of the two-body distribution 
$\widetilde f_2$ underlying Boltzmann ansatz is not an exact
property of the distributions obeying hierarchy (\ref{eq:hierarchy}). 
On the other hand, for $d \geq 1$, the molecular chaos exponents are expected
to become exact. This property is illustrated in the next section.

%%%%%%%%%%%%%%%%%%%%%%%%%%%%%%%%%%%%%%%%%%%%%%%%%%%%%%%%%%%%%%%%%%%%%%%%%%%%%%
\section{Simulation results}
\label{sec:numerical}

The numerical results presented in this section correspond
to the situation $\mu=0$, unless stated. We refer to \cite{Trizac}
for the case of diverging ($\mu<0$) or depleted ($\mu>0$) velocity 
distributions near $v=0$.

%%%%%%%%%%%%%%%%%%%%%%%%%%%%%%%%%%%%%%%%%%%%%%%%%%%%%%%
\subsection{The numerical methods}

We follow two complementary numerical routes. First, we solve
the time-dependent homogeneous Boltzmann equation by means of the
Direct Simulation Monte Carlo method (DSMC), originally developed
to study ordinary gases \cite{Bird}. This scheme, where a suitable
Markov chain is constructed, has been extended to deal with inelastic 
collisions \cite{Montanero,Frezzotti} and is easily modified to describe the
situation under study here, which does not conserve the total number
of particles. Restricting to a spatially homogeneous system, the algorithm
is especially easy to implement, and may be summarized as follows:
among $N_0$ initial particles having a given velocity distribution, 
a pair $(i,j)$ is chosen at random, and removed from the system with a probability
proportional to $|{\bf v}_{ij}|$. The (suitably renormalized) time variable is then
incremented by the amount $(N^2 |{\bf v}_{ij}|)^{-1}$, where $N$ is the number
of particles remaining in the system, before another pair is drawn etc\ldots
This scheme provides the numerical exact solution of Eq. (\ref{eq:boltznu}),
and allows to test the validity of the approximations put forward
in section \ref{sec:Boltzmann} [essentially: Truncation at second order
in expansion (\ref{eq:Sonine}) supplemented with calculations performed
at linear order in the fourth cumulant $a_2$, see Appendix \ref{app:Sonine}]. 
A precise analysis of the late time dynamics (and especially the 
computation of velocity distributions) suffers from the 
concomitant low number of particles left, and the statistical accuracy
is improved by averaging over independent realizations.

The second numerical method (Molecular Dynamics \cite{Allen}) 
consists of integrating the exact equations
of motion for an assembly of spheres confined in a (hyper)-cubic
box with periodic boundary conditions. This route assesses the validity 
of the approach relying on the homogeneous Boltzmann equation, but does not offer
the same accuracy as DSMC, nor the possibility to follow the evolution
over comparable times. In particular, once the mean-free-path $\lambda$, 
which grows
rapidly as $t^\xi$, becomes of the order of the box size $L$, the subsequent
evolution suffers from finite size effects and should be discarded: 
When $\lambda > L$, the algorithm used is unable to find collision events for
those particles which make more than one free flight round on the torus 
topologically equivalent to the simulation box, which causes a spurious
slowing down of the dynamics. It is then tempting to reduce $\lambda$ by increasing
particles diameter $\sigma$, but then finite (and necessarily transient) density
effects --incompatible with the scaling assumption (\ref{eq:scaling2})-- 
may also arise if the packing fraction $\phi$, proportional
to $n \sigma^d$, is not low enough. 
Simulating explicitly the limit of point particles,  
the DSMC scheme considered here is free of this defect.
The initial number of particles considered in MD needs to be large 
to allow the system to enter the scaling regime before finite-size 
effects become dominant:
we considered systems with $N=5.10^4$ to $N=5.10^5$ spheres initially
(compared to $N=10^6$ to $10^8$ in DSMC). 

%%%%%%%%%%%%%%%%%%%%%%%%%%%%%%%%%%%%%%%%%%%%%%%%%%%%%%%
\subsection{Dynamic scaling behaviour}

The results of two-dimensional DSMC and MD simulations are shown in
Figure \ref{fig:testxi}, where the quantity on the $x$ axis is expected to 
scale like real time $t$ from the scaling relation $\xi+\gamma=1$. 
This log-log plot is a direct probe of the exponent $\xi$,
from the slope measured. Both MD and DSMC methods give compatible results,
with the possibility to follow the dynamics over a longer time interval
in DSMC. The departure observed for $n_0/n \simeq 200$ corresponds to the
slowing down of MD dynamics resulting from finite-size effects (see Figure
\ref{fig:unsurndet} below). The theoretical predictions at zeroth
and second order are very close ($\xi_0 = 0.872$ and $\xi_2=0.870$),
and in excellent agreement with the simulation results over several decades.
On a similar graph as Fig. \ref{fig:testxi}, the kinetic ``temperature'' 
$\overline v^2$ exhibits a lower law behaviour (not shown)
with an exponent $-2 \gamma$ in good agreement with the theoretical 
prediction ($\gamma \simeq 0.13$ in 2D). Moreover, the exponents obtained 
analytically and numerically are compatible to those reported in 
the literature ($\xi \simeq 0.89$ in \cite{BenNaim1} and 
$\xi \simeq 0.87(5)$ using a multi-particle lattice gas method
\cite{Chopard}). 

The time evolution of inverse density and inverse typical velocity square 
is shown in Figure \ref{fig:unsurndet}, where considering $n_0/n-1$ and 
$\overline v_0^2/\overline v^2-1$ instead of $n_0/n$ and 
$\overline v_0^2/\overline v^2$ allows to probe the short time behaviour.
Unless stated, the initial velocity distribution is an isotropic Gaussian.
From Eqs. (\ref{eq:alphadef1}) and  (\ref{eq:alphadef2}), $n$ and 
$\overline v^2$ evolve linearly with $t$ for $\omega_0 t \ll 1$ 
[see also (\ref{eq:timen}) and (\ref{eq:timev})]; the same
holds for inverse density and inverse typical velocity squared, 
which is indeed observed in Fig. \ref{fig:unsurndet}.
MD and DSMC result super-impose, except at late times where MD dynamics
suffers from the slowing down discussed previously. 
For both numerical methods, the scaling relation $\xi+\gamma=1$
is well obeyed, in principle at late times only, in the scaling regime.
Special combinations of $n$ and $\overline v$ can however be constructed
with the requirement to match the short time evolution with the scaling behaviour. 
One of these quantities is displayed in Fig. \ref{fig:scaling}, with a 
resulting scaling regime extending over more than 10 decades in time. 
In Fig. \ref{fig:renorm}, we not only test the validity of the theoretical 
scaling exponents, 
but also the full time dependence as predicted by 
Eqs. (\ref{eq:timen}) and (\ref{eq:timev}). 
In order to improve the agreement between
theory and simulation (which holds over more than 6 orders of magnitude in time), 
the system has been left time to enter the scaling
regime: The time origin $t=0$ has been chosen when 80\% of the 
particles originally present have disappeared. 
The corresponding reference configuration (with subscripts 0) 
thus differs from the ones considered previously.

%%%%%%%%%%%%%%%%%%%%%%%%%%%%%%%%%%%%%%%%%%%%%%%%%%%%%%%
\subsection{Velocity distribution in the scaling regime}

In order to understand the reasons for the good agreement between 
our theoretical predictions and the simulations, we now consider the
velocity distribution, restricting to Monte Carlo results
(leading to similar conclusions, 
MD is much more demanding on CPU time and does not allow 
to investigate detailed features of \smash{$\widetilde f_1$} 
with the same accuracy). After a transient where the probability 
distribution function $\widetilde f_1$ evolves with time, 
a well defined scaling regime is reached with a time independent
$\widetilde f_1({\bf c})$ shown in Fig. \ref{fig:pdf} together
with the Sonine prediction pushed at second order. The agreement is 
remarkable, and it is also observed that the Gaussian approximation 
is already close to the asymptotic rescaled velocity distribution,
which is quite surprising in a kinetic process extremely far from equilibrium,
with furthermore no conservation laws.
Given that our perturbative analytical work relies on the calculation
on low order moments of $\widetilde f_1$, this explains the  
accuracy of the zeroth order predictions $\alpha_0$ and $\xi_0$. 
From Eq. (\ref{eq:tail}), we expect the differences between the Sonine 
expansion and the
numerical distribution to become visible in the high energy tail, 
which is confirmed by Fig. \ref{fig:pdf_tail}. As predicted, $\widetilde f_1$
is overpopulated with respect to the Gaussian, and displays a high velocity tail
of the form (\ref{eq:tail}) (see the inset).

%%%%%%%%%%%%%%%%%%%%%%%%%%%%%%%%%%%%%%%%%%%%%%%%%%%%%%%
\subsection{Evolution toward the asymptotic solutions}

Before the scaling regime is attained, $\widetilde f_1$ is time-dependent,
as shown in Fig. \ref{fig:sonine_max}, where the distributions at different
times have been renormalized by ${\cal M}$ to emphasize the building-up of
non-Gaussianities. The evolution towards the scaling solution $1+a_2 S_2$
can be observed. With respect to the Gaussian, $\widetilde f_1$ is at all times
overpopulated both at large and small velocities (which may be related
to the positive sign of $a_2$ for the latter case); normalization is ensured
by an under-population at intermediate velocities. Figures \ref{fig:pdf},
\ref{fig:pdf_tail} and \ref{fig:sonine_max} show that the indirect measure
of $a_2$ through the non-Gaussian character of $\widetilde f_1/{\cal M}$ 
agrees with the theoretical prediction, but it is also possible to 
compute directly $a_2$ in the simulations through its definition as
a fourth cumulant [Eq. (\ref{eq:a2def})]. It turns that both methods
are numerically fully compatible. Moreover, the Sonine expansion 
(\ref{eq:Sonine}) truncated at $n=2$ holds at any time, even in the
transient regime, with the time dependent fourth cumulant $a_2$ 
measured from (\ref{eq:a2def}) (see Fig. \ref{fig:max_transient}).
This result is not a priori expected and points to the relevance of the
expansion (\ref{eq:Sonine}). We did not try to solve analytically
the homogeneous time-dependent Boltzmann equation within the same framework as in the
scaling regime, so that we do not have any prediction for the (transient) 
time-dependence of $a_2$ and $\alpha$. However, as shown in the inset of 
Fig. \ref{fig:max_transient}, the relation (\ref{eq:apptransient})
(which reads in 2D $\alpha = 5/4 + 7 a_2/16$) remarkably holds for all time.
Here, the energy dissipation parameter $\alpha$ has been computed
through the ratio 
$\langle c_{12}\,c_1^2\rangle/(\langle c_{12}\rangle\langle c_1^2\rangle) = 
\langle c_1^2\rangle_{\hbox{\scriptsize coll}}/\langle c_1^2\rangle$,
where $\langle c_1^2\rangle_{\hbox{\scriptsize coll}}$, the mean energy dissipated 
in a collision is computed in the simulations and normalized by the
time-dependent mean kinetic energy per particle $\langle c_1^2\rangle$.

%%%%%%%%%%%%%%%%%%%%%%%%%%%%%%%%%%%%%%%%%%%%%%%%%%%%%%%
\subsection{A final remark: Identification of ``isobestic'' points}

For $\mu=0$, Figure \ref{fig:sonine_max}  indicates that 
during the transient evolution toward scaling,
the distributions of reduced velocities have fixed points (for a given initial 
velocity distribution, the curves corresponding to $\widetilde f_1$
at different times all pass through common points,
that we called isobestic points).  
This feature has been observed for 
all initial distributions investigated and appears to be a systematic property
of the dynamics, which still holds for non vanishing values of $\mu$ (see Fig.
\ref{fig:isobestic}). We have no analytical explanation for this observation.

%%%%%%%%%%%%%%%%%%%%%%%%%%%%%%%%%%%%%%%%%%%%%%%%%%%%%%%%%%%%%%%%%%%%%%%%%%%%%
\section{Conclusion}
\label{sec:concl}

An analytical derivation of the equations governing the dynamics of
an infinite system of  spherical particles in a $d$-dimensional
space, moving freely between collisions and annihilating in pairs when
meeting, has been obtained.
The hierarchy equations obeyed by the reduced distributions 
$f_k(1,2,..., k;t)$  have been derived.  In the Grad limit, this hierarchy 
formally reduces
to a  Boltzmann-like hierarchy for dimensions $d>1$.
If these reduced distributions $f^B_k$ factorize at the initial time, this 
factorization remains valid for all times and the whole hierarchy reduces to one
non-linear equation.

In the long time limit, the ratio of the particle radius to 
the mean-free path vanishes.
A scaling analysis of the exact homogeneous hierarchy equations has been performed.
Self-similar reduced distributions in which the time dependence has been absorbed into
the density $n(t)$ and the root mean-square velocity ${\overline v}(t)$ were introduced. 
As a result, the exponents $\xi$ and $\gamma$ describing the decay  with time 
of  $n(t)$ and ${\overline v}(t)$
depend only upon one single parameter $\alpha$, related to the dissipation of energy.
Moreover, it turns out that in dimension higher than $1$, 
the terms responsible for the violation of molecular chaos
are asymptotically irrelevant. Therefore, we recover the conclusions 
reached in the formal
Grad limit and thus, the Boltzmann equation becomes exact 
in the long time limit in dimensions higher than $1$. 
The above arguments give a first principle justification for the use
of the Boltzmann equation approach for $d>1$, as well as its limitations, 
in ballistic annihilation problems, an issue that has been overlooked so far.
However, as discussed above, this scaling analysis is incorrect for a $1$
dimensional system with discrete initial velocity distribution,
a situation for which the scaling assumption underlying our approach fails.

The Boltzmann equation has been solved within an 
expansion in Sonine polynomials $S_n$. Truncation to order
$n=2$, provides the first non-Gaussian
corrections to the scaled velocity distribution, and
leads to analytical predictions for the exponents $\xi$ and
$\gamma$ as a function of $d$. 
For large dimension $d$, these predictions coincide with the exact
$1/d$ correction to the naive mean field values
($\xi=1$ and $\gamma=0$), calculated in \cite{Krapivsky}.
The above analytical predictions are in remarkable agreement with the results
of extensive numerical simulations we have performed for two dimensional systems
(implementing the complementary Monte Carlo and Molecular Dynamics techniques). 
In 1D for regular continuous velocity distributions, 
it is noteworthy that they are in excellent agreement with the
numerical solution of the Boltzmann equation, and quite close to the
exact values obtained with Molecular Dynamics (4\% difference). 
This last point was unexpected since molecular
chaos breaks in 1D. Finally, the time dependence of the reduced velocity distribution
function shows an unexplained and remarkable feature, with the existence 
of fixed (``isobestic'') points, irrespective of initial conditions.

\vskip 0.5cm 
Acknowledgments: We would like to thank F. Coppex for a careful reading
of the manuscript, T. van Noije, A. Barrat, F. van Wijland and
A. Santos for useful discussions. E.T. acknowledges the hospitality
of the Theoretical Physics Department (Gen\`eve) where part of this work
was performed. J.P. and E.T. benefited from the financial support of the
``Programme d'Action Integr\'ee Franco-Polonais Polonium''
(contract 03990WF). M.D. acknowledges the support of the Swiss National Science 
Foundation and of the CNRS through the attribution of a 
``poste de chercheur associ\'e''.

%%%%%%%%%%%%%%%%%%%%%%%%%%%%%%%%%%%%%%%%%%%%%%%%%%%%%%%%%%%%%%%%%%%%%%%%%%%%%
\appendix

\section{}
\label{app:alpha}

Within Boltzmann's kinetic equation, we show in this appendix that 
$\alpha>1$. To this end, Eq. 
(\ref{eq:Boltzrescaled}) is rewritten in the form
\be
\widetilde f_1( c_1) + 
\left(\frac{1-\alpha}{2}\right)
\hbox{div}_{{\bm c}_1}\left({\bm c_1} \widetilde f_1( c_1)\right) \,
\,=\,
\widetilde f_1(c_1) \int\! d{\bm c_2}\, 
\frac{c_{12}}{\langle c_{12}\rangle}\, \widetilde f_1 (c_2),
\label{eq:Boltzapp}
\ee
and integrated with weight $\psi(c_1) \,d{\bf c}_1$, where we choose
$\psi(c) = c^{2/(1-\alpha)}$. Assuming that 
the moment \smash{$\int \psi \widetilde f_1$} exists, 
we obtain after 
an integration by parts (neglecting surface terms)
\be
0 = \int\! d{\bm c}_1 d{\bm c}_2 \,\psi(c_1) \,
\frac{c_{12}}{\langle c_{12}\rangle}\, \widetilde f_1 (c_1)
\,\widetilde f_1 (c_2).
\label{eq:appaa}
\ee
The right hand side of Eq. (\ref{eq:appaa}) is a strictly positive 
quantity [except when \smash{$\widetilde f_1 (c)=\delta(c)$}], which 
leads to a contradiction. We therefore conclude that the 
quantity $\int \psi \widetilde f_1$ does not exist. Remembering that
$\psi(c) \widetilde f_1(c) \sim c^{\mu+2/(1-\alpha)}$ near the velocity
origin, the divergence of $\int \psi \widetilde f_1$ implies
\be
\frac{2}{1-\alpha} +d -1 + \mu < -1.
\ee
Supplementing this condition with the normalization constraint $\mu >-d$, 
we obtain $\alpha>1$. The physical meaning of this condition is that
the typical kinetic energy dissipated per particle in a collision 
is larger than the average kinetic energy of the system at the same time.
The temperature $\overline v^2$ is therefore a decreasing function of time.

%%%%%%%%%%%%%%%%%%%%%%%%%%%%%%%%%%%%%%%%%%%%%%%%%%%%%%
\section{Second order truncated Sonine expansion}
\label{app:Sonine}
In this Appendix, we calculate the dominant deviation of $\widetilde f_1$
from the Gaussian shape, and the associated energy dissipation parameter $\alpha$
by setting $\widetilde f_1(c) = {\cal M}(c)\left\{1+ a_2 S_2(c^2)
\right\}$. The moments appearing in (\ref{eq:alphautile0}) 
and (\ref{eq:alphautile1}) with $p=2$ are then computed as a function of $a_2$. 
Upon writing 
\be
\alpha \,=\,  1+ \frac{2}{\mu + d}\left(
1-\frac{\langle c_1 \rangle}{\langle c_{12}\rangle}\right) \,=\,
\frac{\langle c_{12}\,c_1^2\rangle}{\langle c_{12}\rangle\langle c_1^2\rangle},
\label{eq:app1}
\ee
the last equality provides an equation for $a_2$ which is solved 
so that $\alpha$ is finally explicitly known as a function of input
parameters $\mu$ and dimension $d$. As in the main text, the angular brackets
denote averages with weight $\widetilde f_1(c_1) \widetilde f_1(c_2) $
\begin{eqnarray}
\langle \ldots \rangle  &=&  \int (\ldots) \, \widetilde f_1(c_1) \widetilde f_1(c_2)\,
d{\bf c}_1 d{\bf c}_2 \\
&=& \int (\ldots)\, {\cal M}(c_1) {\cal M}(c_2) \left\{
1 + a_2 \left[ S_2(c_1^2) + S_2(c_2^2)
\right] \right\}\, d{\bf c}_1 d{\bf c}_2 \,+\, {\cal O}\left( a_2^2\right).
\end{eqnarray}
In the following, non-linear terms of order $a_2^2$ will be neglected. 
In the spirit of \cite{Pre2}, 
it is convenient to introduce center-of-mass and relative velocities
${\bf c}_1 = {\bf C} + {\bf c}_{12}/2$; ${\bf c}_2 = {\bf C} - {\bf c}_{12}/2$ 
and to define the generic moments
\begin{eqnarray}
M_{np} & = & \left\langle c_{12}^n \,C^p \,\right\rangle \label{eq:appmnp}\\
  & = & \int \!d {\bf c}_{12} d{\bf C} \, c_{12}^n \,\,C^p
  \left(\frac{d}{2\pi}\right)^{d} \, e^{-d c_{12}^2/4 -  d C^2} \,
  \left\{ 1 + a_2 \left[\frac{d^2}{8} \left(c_1^4 + c_2^4 \right) -  
  \frac{d(d+2)}{4} \left(c_1^2 + c_2^2 \right) + \frac{d(d+2)}{4}
  \right]\right\} 
  \label{eq:soninint}
\end{eqnarray}
From $c_1^4 + c_2^4 = 2 C^4 + 2({\bf C}\cdot {\bf c}_{12})^2 + 
c_{12}^4/8 + C^2 c_{12}^2$, the term $({\bf C}\cdot {\bf c}_{12})^2$ appearing 
under the integral sign in (\ref{eq:soninint}) becomes $C^2 c_{12}^2/d$ 
for symmetry reasons, and making use of $c_1^2 + c_2^2 = 2C^2 + c_{12}^2/2$, 
the variables ${\bf c}_{12}$ and ${\bf C}$ decouple in (\ref{eq:soninint}). 
The resulting integrals yield
\be
\frac{M_{np}}{M_{np}^0} \,=\, \frac{\left\langle c_{12}^n \,C^p \,
\right\rangle_{~}}{\left\langle c_{12}^n \,C^p \,\right\rangle_0} \,=\,
1 + \frac{a_2}{16 d} \left\{d(n^2+p^2) - 2 d (n+p) + 2 n p (d+2)
\right\} \,+\,{\cal O}\left( a_2^2 \right).
\label{eq:appmres}
\ee
In this equation, the subscript $0$ refers to averages with Gaussian measure
(formally $a_2=0$):
\be
\left\langle c_{12}^n \,C^p \,\right\rangle_0  =  
(\sqrt{d})^{-n-p} \, 2^n\, \frac{\Gamma\left(\frac{d+n}{2}\right)\,
\Gamma\left(\frac{d+p}{2}\right)}{\Gamma^2(d/2)},
\ee 
where $\Gamma$ is the Euler function.
The moments $\langle c_{12}\rangle$  and $\langle c_{12} \,c_1^2\rangle = 
\langle c_{12} (c_1^2+c_2^2)\rangle/2 = \langle c^3_{12}\rangle /4  
+ \langle c_{12}\, C^2\rangle$ appearing in (\ref{eq:app1})
are then known 
\begin{eqnarray}
&& \frac{\left\langle c_{12} \right\rangle_{~}}{\left\langle c_{12}\right\rangle_0} 
\,=\, 1-\frac{1}{16} \, a_2  \,+\,{\cal O}\left( a_2^2 \right)\label{eq:appc}\\
&&\frac{\left\langle c_{12}\, c_1^2 \,\right\rangle_{~}}
{\left\langle c_{12}\right\rangle_0} \,=\, 1 + \frac{1}{2d} + \frac{a_2}{32}
\left(6 + \frac{11}{d}\right) \,+\,{\cal O}\left( a_2^2 \right)\\
\Longrightarrow && 
\frac{\langle c_{12}\,c_1^2\rangle}{\langle c_{12}\rangle\langle c_1^2\rangle} \,=\,
1 + \frac{1}{2d} + \frac{a_2}{8} \left( 2 + \frac{3}{d}\right) \,+\,
{\cal O}\left( a_2^2 \right).
\label{eq:apptransient}
\end{eqnarray}
For an elastic hard sphere fluid at equilibrium (with thus $a_2=0$), this
last quantity equals $1+ 1/(2d)$ and represents the ratio of the mean kinetic energy 
of colliding particles (averaged over successive collision events) to the
mean kinetic energy of the population. As expected, this ratio exceeds 1,
since typical colliding partners are ``hotter'' than the mean background.
This quantity is easily measured in Molecular Dynamics or Monte Carlo 
simulations
(see e.g. Fig. \ref{fig:max_transient}). 
We also note that the ratio (\ref{eq:appc}) has been computed in 
\cite{vanNoije} at the same level of approximation in the context of
rapid granular flows, with the same result. van Noije and Ernst
also reported non Gaussian corrections to the cooling rate $\gamma$
of an inelastic hard sphere fluid \cite{vanNoije}
\be
\frac{\gamma}{\gamma_0} \,=\, 1 + \frac{3}{16} a_2 
 \,+\,{\cal O}\left( a_2^2 \right) \qquad \forall d,
\label{eq:appcooling}
\ee
where $\gamma_0$ denotes the cooling rate evaluated assuming Maxwellian
velocity distributions. 
In terms of the moments $M_{np}$ introduced in (\ref{eq:appmnp}), it can be shown
that $\gamma/\gamma_0 = M_{30}/M_{30}^0$ and it is then easily checked
that expression (\ref{eq:appmres}) reduces to (\ref{eq:appcooling}) 
for $n=3$ and $p=0$, for arbitrary dimensionality.

The remaining unknown quantity is $\left\langle c_{1} \right\rangle$ which may
be calculated following similar lines as above:
\begin{eqnarray}
&& \frac{\left\langle c^n_{1} \right\rangle}
{\left\langle c^n_{1} \right\rangle_0} \,=\, 1 + \frac{a_2}{8}n (n-2) 
 \,+\,{\cal O}\left( a_2^2 \right)\\
&& \left\langle c^n_{1} \right\rangle_0 \,=\,
\left(\frac{2}{d}\right)^{n/2}\,\frac{\Gamma\left(\frac{d+n}{2}\right)}{\Gamma(\frac{d}{2})},
\label{eq:appmom}
\end{eqnarray}
from which we extract $\left\langle c_{1} \right\rangle = 
\left\langle c \right\rangle$. 
As expected, the $a_2$ correction in (\ref{eq:appmom}) vanishes for 
$n=0$ and $n=2$, which follows respectively from the normalization 
constraint and the definition (\ref{eq:defrescaled}) 
of ${\bf c}$ implying $\langle c^2 \rangle=1$. Gathering results, we obtain 
(\ref{eq:a2}) and (\ref{eq:alpha2}) from Eqs. (\ref{eq:app1}).

%%%%%%%%%%%%%%%%%%%%%%%%%%%%%%%%%%%%%%%%%%%%%%%%%%%%%%%%%%%%%%%%%%%%%%

%%%%%%%%%%%%%%%%%%%%%%%%%%%%%%%%%%%%%%%%%%%%%%%%%%%%%%%%%%%%%%%%%%%%%%%%%%%%%%%
%%%%%%%%%%%%%%%%%               FIGURES            %%%%%%%%%%%%%%%%%%%%%%%%%%%%
%%%%%%%%%%%%%%%%%%%%%%%%%%%%%%%%%%%%%%%%%%%%%%%%%%%%%%%%%%%%%%%%%%%%%%%%%%%%%%%
%\newpage

\begin{center}
\begin{figure}[htb]
\vspace{4cm}
\epsfig{figure=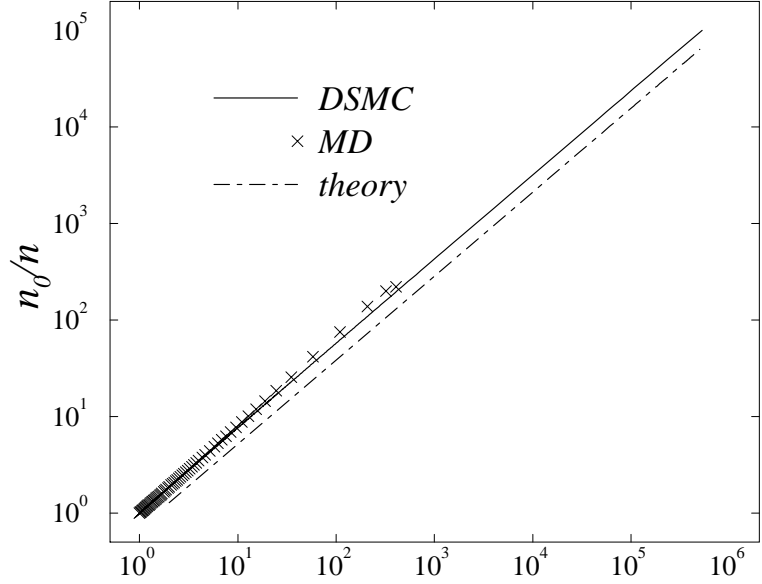,width=10cm,angle=0}
\caption{Inverse density $n_0/n$ versus $n_0 \overline v_0/(n\overline v)$,
for $d=2$. The dotted line has slope $\xi_2 = 0.87$ 
[see equation (\ref{eq:xi0})]. Initial number of particles:
$5.10^6$ (with a further average over $10^3$ independent replicas) for DSMC 
and $2.10^5$ for MD. In both cases, the initial velocity distribution is
Gaussian ($\mu=0$), and the initial configuration used for MD
is that of an equilibrium hard disk fluid with packing fraction $\phi = 0.1$
(chosen low enough to avoid finite packing effects).
\label{fig:testxi}}
\end{figure}
\end{center}

\begin{center}
\begin{figure}[htb]
%\vspace{-0.5cm}
\epsfig{figure=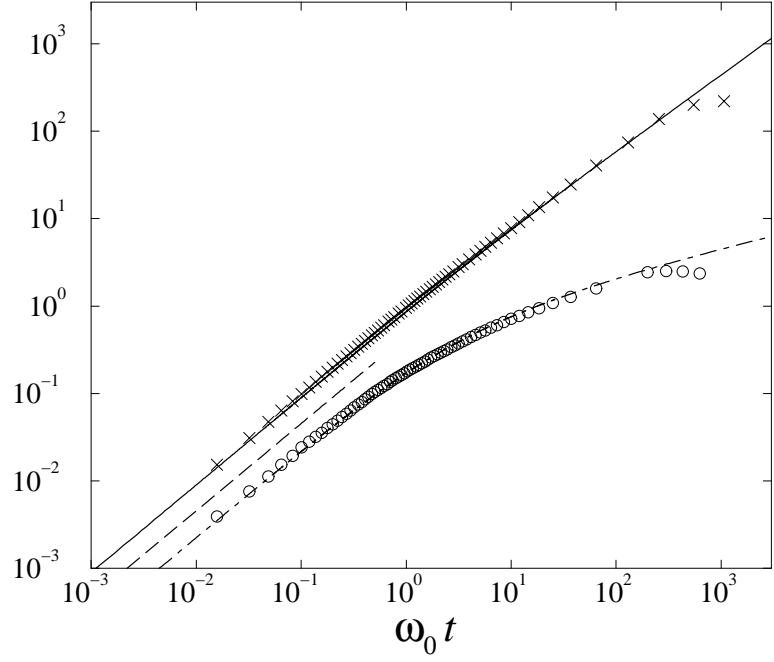,width=10cm,angle=0}
\caption{Plots of $(n_0/n)-1$ [upper curve corresponding to 
DSMC, compared to its MD counterpart (crosses)]
and $(\overline v_0/\overline v)^2-1$
(lower dashed curve for DSMC, circles for MD),
as a function of time. 
The dotted line at short times has slope 1. 
\label{fig:unsurndet}}
\end{figure}
\end{center}

\begin{center}
\begin{figure}[htb]
%\vspace{-0.5cm}
\epsfig{figure=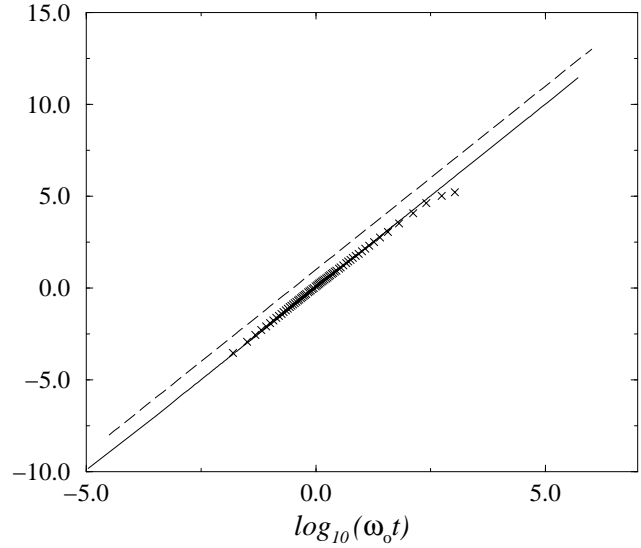,width=9cm,angle=0}
\caption{Plot of 
$\log_{10}[n_0/n-1] + \log_{10}\left[n_0\overline v_0^2/(n\overline v^2)-1 \right]$
on a logarithmic time scale ($d=2$). DSMC results are represented by the continuous
curve, while the crosses correspond to MD. The dashed line has slope 2.
At late times, the quantity displayed is expected to 
behave as $2\log_{10}(\omega_0 t)$ from the scaling relation $\xi +\gamma=1$. 
At short times, the same behaviour is observed, for a different reason
[see Eqs. (\ref{eq:timen}) and (\ref{eq:timev})]. The ultimate MD
slowing down is again visible.
\label{fig:scaling}}
\end{figure}
\end{center}

%\vspace{1cm}

\begin{center}
\begin{figure}[htb]
%\vspace{-0.5cm}
\epsfig{figure=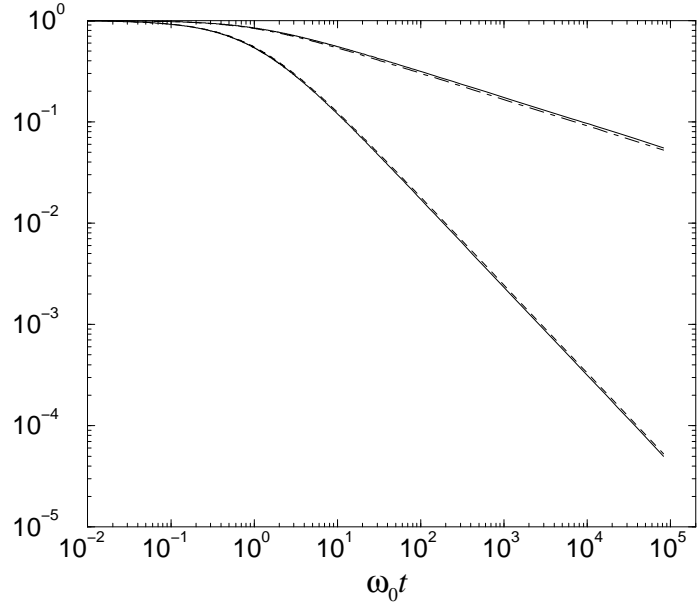,width=9cm,angle=0}
\caption{Time dependence of $n$ (lower curve) and $\overline v^2$
(upper curve) obtained in Monte Carlo, compared to the dashed curves
corresponding to the theoretical
predictions (\ref{eq:timen}) and (\ref{eq:timev}), where the energy
dissipation coefficient $\alpha$ is calculated at second order
in Sonine expansion [$\alpha_2 \simeq 1.297 $ from Eq. (\ref{eq:alpha2})].
\label{fig:renorm}}
\end{figure}
\end{center}

\begin{center}
\begin{figure}[htb]
%\vspace{-0.5cm}
\epsfig{figure=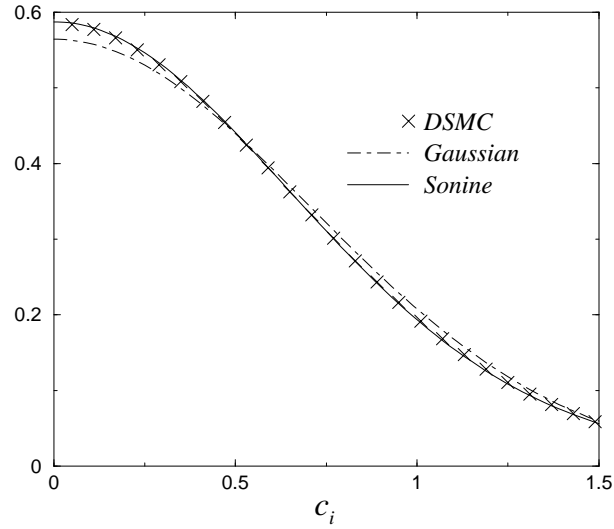,width=8cm,angle=0}
\caption{Probability distribution function $\widetilde f_1(c_i)$ 
of a given Cartesian component $c_i$ of the rescaled 
two-dimensional velocity ${\bf c}$. 
The time independent distribution obtained in DSMC simulations
at late times is compared to the Gaussian ${\cal M}$ and the Sonine expansion
truncated at $n=2$, with the fourth cumulant given by Eq. (\ref{eq:a2})
($a_2 \simeq 0.109$ for $d=2$ and $\mu=0$). All distributions have variance 1/2
so that $\langle c^2\rangle =1$. The results have been obtained by averaging 
over 50 replicas of a system with initially $N=40.10^6$ particles. 
\label{fig:pdf}}
\end{figure}
\end{center}

%\vspace{3cm}

\begin{center}
\begin{figure}[htb]
%\vspace{-0.5cm}
\epsfig{figure=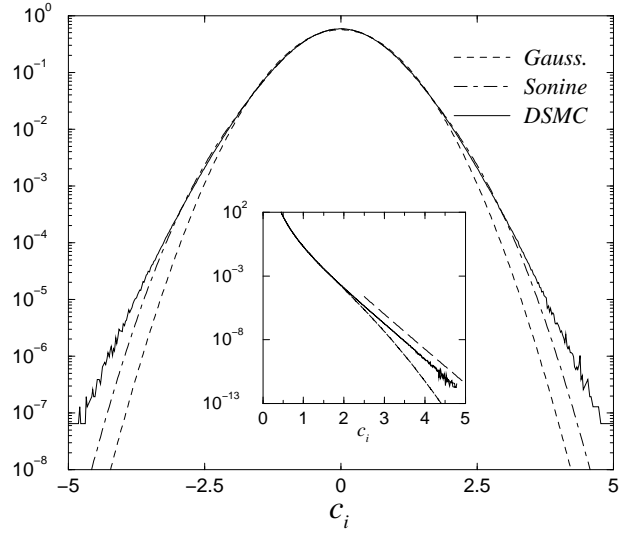,width=8cm,angle=0}
\caption{Same as Figure \ref{fig:pdf}, on a linear-log scale to probe the tails
of the distributions. In the inset showing $c_i^{-4.7} \widetilde f_1(c_i)$ 
as a function of $c_i$, the dashed curve corresponds to the Gaussian 
[$c_i^{-4.7} {\cal M}(c_i)$] while the straight line is a guide for the eye 
evidencing a pure exponential behaviour. For $d=2$ and $\mu=0$, the exponent 
$d + 2/(1-\alpha)$ appearing in Eq. (\ref{eq:tail}) is close to -4.7 
and has been used to rescale the quantity plotted on the $y$-axis in the inset. 
\label{fig:pdf_tail}}
\end{figure}
\end{center}

\begin{center}
\begin{figure}[htb]
%\vspace{-0.5cm}
\epsfig{figure=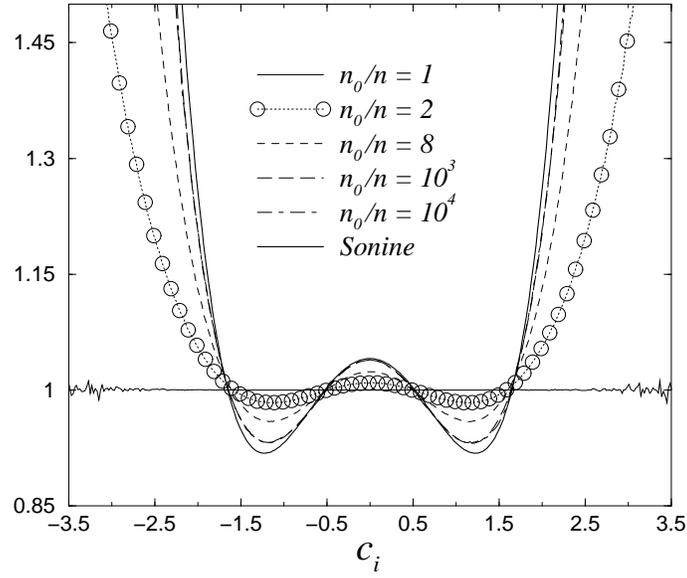,width=9cm,angle=0}
\caption{Plots of $\widetilde f(c_i)/{\cal M}(c_i)$ versus $c_i$,
at different times corresponding to the indicated densities.
The initial distribution is Gaussian (thus corresponding to the flat
curve), and the thick curve is Sonine solution $1+a_2 S_2$
with $a_2$ given by (\ref{eq:a2}).
\label{fig:sonine_max}}
\end{figure}
\end{center}

\begin{center}
\begin{figure}[htb]
%\vspace{-0.5cm}
\epsfig{figure=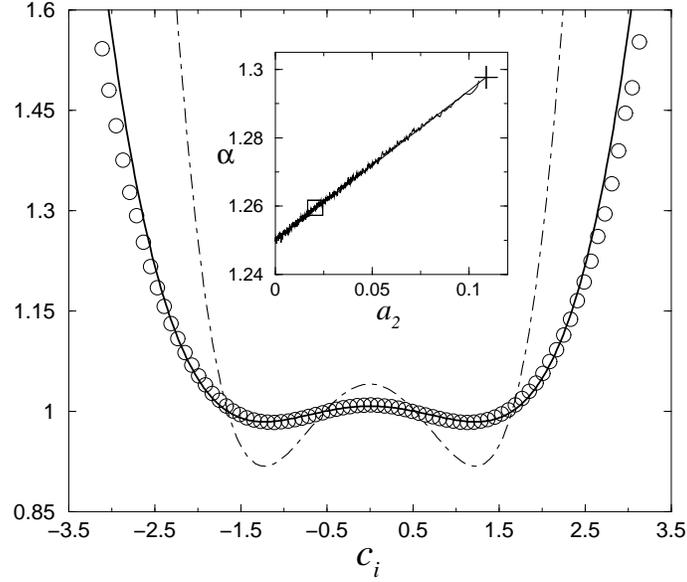,width=9cm,angle=0}
\caption{Plot of $\widetilde f(c_i)/{\cal M}(c_i)$ as a function of $c_i$,
at the particular time $t_{1/2}$ where the density is exactly 
half the initial one,
with the same initial distribution as in Fig. \ref{fig:sonine_max}
[the circles (DSMC measure) thus show the same distribution as the circles of  
Fig. \ref{fig:sonine_max}]. The thick curve shows $1+a_2(t_{1/2}) S_2$
with $a_2(t_{1/2})$ measured from its definition (\ref{eq:a2def}).
The dashed curve is the Sonine prediction in the scaling regime
(i.e. the thick curve of Fig. \ref{fig:sonine_max}).
Inset: $\alpha$ as a function of $a_2$ (see main text). The DSMC measure
is compared to the prediction (\ref{eq:apptransient}) shown by the straight
line ending at the point --indicated by a cross--
of coordinates (0.109,1.297) as predicted
by Eqs. (\ref{eq:a2}) and (\ref{eq:alpha2}). The square located at (0.0207,1.2595)
corresponds to the numerical measure of $\alpha$ and $a_2$
made at time $t_{1/2}$ for which the velocity 
distribution is displayed  in the main graph.
\label{fig:max_transient}}
\end{figure}
\end{center}

\begin{center}
\begin{figure}[htb]
%\vspace{-0.5cm}
\epsfig{figure=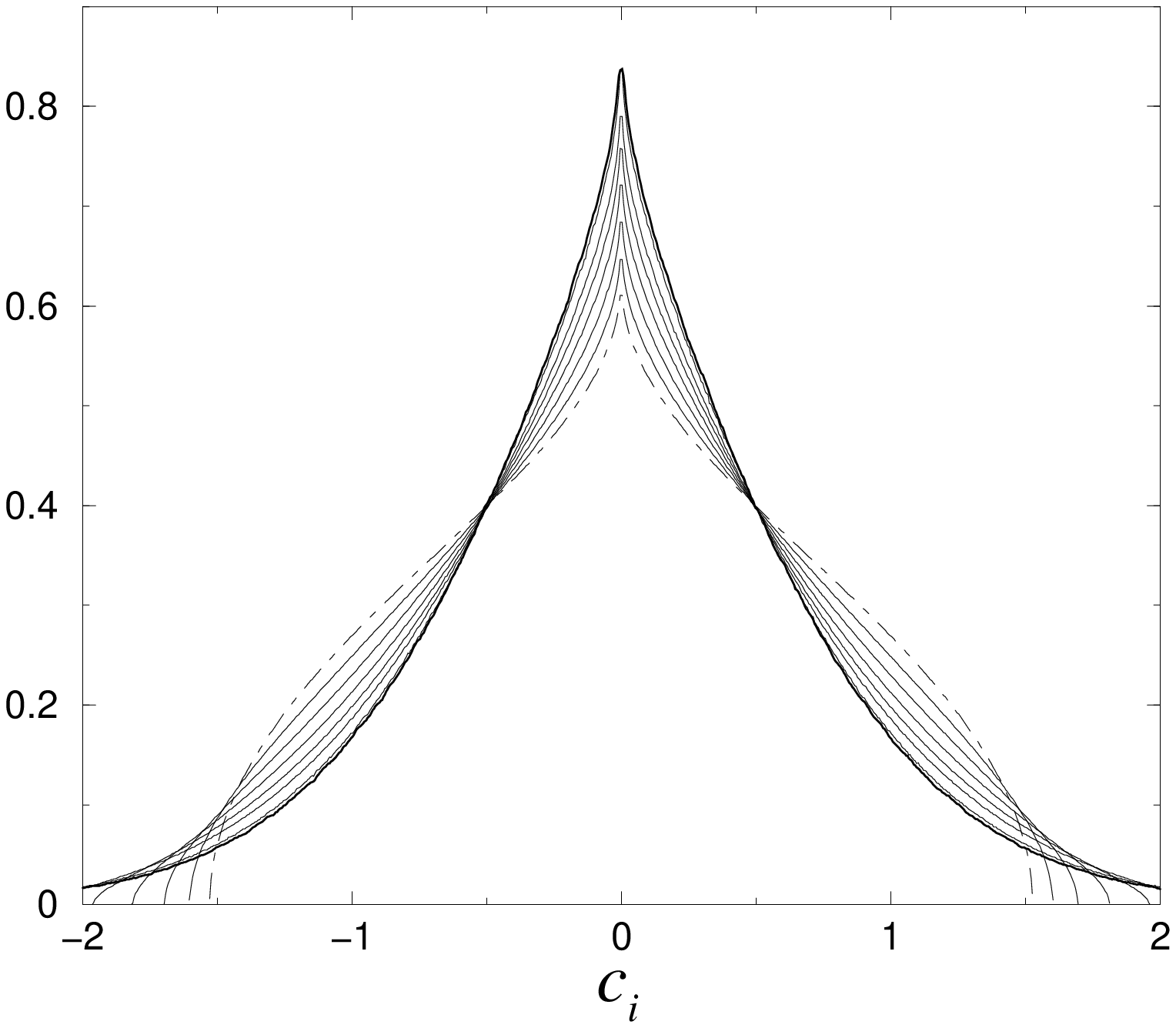,width=8cm,angle=0}
\hspace{0.2cm}
\epsfig{figure=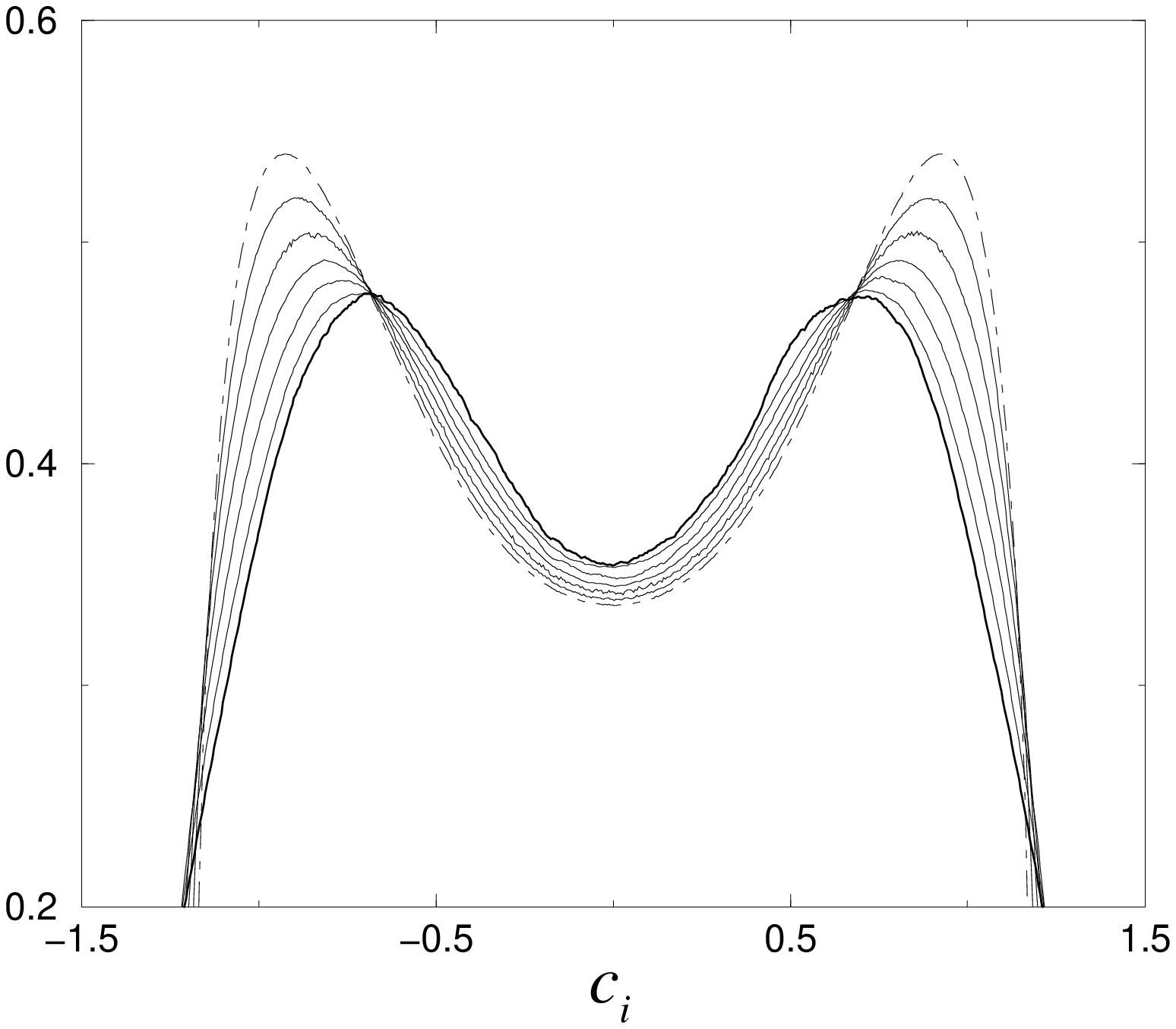,width=8cm,angle=0}
\caption{Plots of $\widetilde f_1(c_i)$ as a function of $c_i$
at different times. The left graph corresponds
to an initial velocity distribution with $\mu=-3/2$
while $\mu=3$ for the right graph. On both graphs, the initial 
distribution is shown by the dashed curve, whereas the thick curves
display the asymptotic distributions approached in the scaling regime. 
%Remark on marginal not vanishing at $c_i=0$. 
\label{fig:isobestic}}
\end{figure}
\end{center}

\end{document}